\documentclass[11pt,twoside]{article}
\usepackage{mathrsfs}
\usepackage{amsmath}
\usepackage{amssymb}
\usepackage{graphicx}
\usepackage{subfigure}
\usepackage{appendix}
\usepackage{cite}
\numberwithin{equation}{section}
\newtheorem{definition}{Definition}

\newtheorem{proposition}{Proposition}
\newtheorem{lemma}{Lemma}

\newtheorem{corollary}{Corollary}
\newtheorem{remark}{Remark}

\DeclareMathOperator{\im}{Im}

\DeclareMathOperator{\tr}{tr}
\DeclareMathOperator{\diag}{diag}
\newcommand*{\QEDB}{\hfill\ensuremath{\square}}

\title{A type I defect and new integrable boundary conditions for the coupled nonlinear Schr\"{o}dinger equation}
\author{
Baoqiang Xia
\\
School of Mathematics and Statistics, Jiangsu Normal University,
\\
Xuzhou, Jiangsu 221116, P. R. China,
\\
E-mail address: xiabaoqiang@126.com
}

\date{}
\begin{document}
\maketitle
\begin{abstract}
We study two integrable systems associated with the coupled NLS equation: the integrable defect system and the integrable boundary systems. Regarding the first one, we present a type I defect condition, which is described by a B\"{a}cklund transformation frozen at the defect location. For the resulting defect system, we prove its integrability both by showing the existence of an infinite set of conserved quantities and by implementing the classical $r$-matrix method. Regarding the second one, we present some new integrable boundary conditions for the coupled NLS equation by imposing suitable reductions on the defect conditions. Our new boundary conditions, unlike the usual boundary conditions (such as the Robin boundary), involve time derivatives of the coupled NLS fields and are characterised by non constant $K(\lambda)$ matrices. We prove the integrability of our new boundary conditions by using Sklyanin's approach.

\noindent {\bf Keywords:}\quad integrable defect problems, integrable boundary conditions, B\"{a}cklund transformation, coupled nonlinear Schr\"{o}dinger equation.

\end{abstract}

\section{ Introduction}

The coupled nonlinear Schr\"{o}dinger (NLS) equation, also known as Manakov equation,
\begin{eqnarray}
\begin{split}
iu_{1,t}+u_{1,xx}+2 \left(|u_1|^2+|u_2|^2\right)u_1=0,
\\
iu_{2,t}+u_{2,xx}+2 \left(|u_1|^2+|u_2|^2\right)u_2=0,
\end{split}
\label{CNLS}
\end{eqnarray}
was introduced by Manakov in \cite{Manakov}.
This equation is a physically and mathematically important integrable model.
It describes some physical phenomena, such as the propagation of an optical pulse in a birefringent optical fiber \cite{Manakov} and the multi-component Bose-Einstein condensates \cite{BuAn}.

In the present paper, we study two integrable systems associated with the coupled NLS equation.
The first one is concerned with the so-called integrable defect problem.
In classical integrable field theories, a defect is introduced as a discontinuity at a specific point together with suitable sewing conditions across the defect relating the fields and their derivatives on either side (see for example \cite{BCZ20041,CZ2006,CZ2009}).
The presence of defects usually spoils the integrability of a system.
The idea of using frozen B\"{a}cklund transformations (BTs) as defect conditions is a significant observation \cite{BCZ20041}. This type of defect has a Lagrangian description, and it preserves the usual properties of classical integrability, such as the existence of (suitably adapted) Lax pairs, the existence of an infinite number of conserved quantities, and the possibility of implementing the method of the classical $r$-matrix; see for example \cite{BCZ20041,CZ2006,HK2008,Caudrelier2008,CK2015,AD2012,Doikou2016}.
So far, there are mainly two types of integrable defect, known as type I and type II defect \cite{BCZ20041,CZ2009}.
The main difference between these two types defect is:
the type I defect has no degrees of freedom, whereas the type II defect requires the
presence of extra degrees of freedom at the defect location.
The type I defect has been well investigated in the context of the NLS equation (see for example \cite{CZ2006,Caudrelier2008,CK2015}).
However, as pointed out in \cite{Zambon2014} (and to our knowledge), no type I-like defects are known for
the coupled NLS equation yet, despite the type II defect problem has been investigated in \cite{Doikou2014}. It is interesting to find out them and then investigate the Liouville integrability of the resulting defect system. This is precisely the first objective of the present paper.
We note that the Lax pair of the coupled NLS equation, in comparison with that of the NLS equation, involves $3\times 3$ matrices, rather than $2\times 2$ ones.
Due to this, it is not easy to find a BT that is suitable for describing the type I defect condition for the coupled NLS equation.
Here we provide such a BT, and we present a type I defect condition for the coupled NLS equation based on this BT
(as in the case of the scalar NLS equation \cite{CZ2006,Caudrelier2008,CK2015}, we define the defect condition by freezing the BT at defect location).
We show that the resulting defect system also admits a Lagrangian description.
We establish the Liouville integrability of the resulting defect system both by showing the existence of an infinite set of conserved quantities and by implementing the classical $r$-matrix method.
In particular, following the analogous argument for the NLS equation \cite{CK2015} (see also \cite{ACDK2016} for some relevant discussion), we introduce the new equal-space Poisson structure for the coupled NLS equation in order to implement the $r$-matrix method to the defect coupled NLS system.
Our results extend the results of \cite{Caudrelier2008,CK2015} from the integrable equations with $2\times 2$ Lax pairs to the one with a $3\times 3$ Lax pair.

The second subject of the present paper is concerned with the integrable boundary conditions associated with the coupled NLS equation on the half-line. By imposing suitable reductions on the defect condition (the BT fixed at $x=0$), we present some new integrable boundary conditions for the coupled NLS equation (see sections 5.1 and 5.2).
These new boundary conditions involve time derivatives of the coupled NLS fields;
they provide two-components generalizations of a new boundary condition presented by Zambon in \cite{Zambon2014} in the context of the NLS equation (see also \cite{Xia2021,Gruner2020,Zhangc2021,CCD2021} for very recent studies on the Zambon's new boundary condition).
By using Sklyanin's approach \cite{Sklyanin1987}, we establish the integrability of our new boundary conditions.
The main technical difficulty in the implementation of Sklyanin's approach to our new boundary conditions is to find out the corresponding boundary $K(\lambda)$ matrices that describe our new boundary conditions.
To overcome this difficulty, we present a connection between the matrices describing the BTs and the boundary $K(\lambda)$ matrices (see proposition \ref{pro2} in section \ref{sec5.2}).
Based on this, we are able to construct explicitly the boundary $K(\lambda)$ matrices for all our new boundary conditions. These boundary $K(\lambda)$ matrices are no more constant matrices; they are characterized by the presence of the coupled NLS fields at the boundary location.
In addition, we also derive generating functions of the conserved quantities for the coupled NLS equation in the presence of our new boundary conditions.

The paper is organized as follows.
As a preparation for the study of the integrability of our defect system,
in section 2 we first briefly review the usual equal-time and the new equal-space Poisson brackets associated with the coupled NLS equation, then we describe the integrability of the coupled NLS equation from
the point of view of the new equal-space Poisson brackets. In section 3, we first derive a type I BT for the coupled NLS equation and then we study its canonical property with respect to both the equal-time and the equal-space Poisson brackets.
In section 4, we present a type I defect based on the BT derived in section 3, and we study the Liouville integrability of the resulting defect system. In section 5, we study a class of new boundary conditions associated with the coupled NLS equation and its Liouville integrability.
Some concluding remarks are drawn in section 6.

\section{Dual Hamiltonian descriptions for the coupled NLS equation}

Before going to the main issue of this section, we first recall that the coupled NLS equation (\ref{CNLS}) is associated to the following Lax pair \cite{Manakov}
\begin{subequations}
\begin{eqnarray}
\phi_x(x,t,\lambda)=U(x,t,\lambda)\phi(x,t,\lambda),
 \label{lpx}
 \\
\phi_t(x,t,\lambda)=V(x,t,\lambda)\phi(x,t,\lambda),
 \label{lpt}
\end{eqnarray}
\label{LPxt}
\end{subequations}
where $\lambda$ is a spectral parameter, $\phi=(\phi_{1},~\phi_{2},~\phi_{3})^T$, and
\begin{subequations}
\begin{eqnarray}
U(x,t,\lambda)&=&\left( \begin{array}{ccc}
-2i\lambda & u_1 & u_2 \\
 -\bar{u}_1 &  i\lambda & 0 \\
 -\bar{u}_2 & 0 & i\lambda \\
 \end{array} \right),
\label{U}
\\
V(x,t,\lambda)&=&\left( \begin{array}{ccc}
-6i\lambda^2+i\left(|u_1|^2+|u_2|^2\right) & 3\lambda u_1+iu_{1,x} &  3\lambda u_2+iu_{2,x} \\
 -3\lambda \bar{u}_1+i\bar{u}_{1,x} & 3i\lambda^2-i|u_1|^2 & -i\bar{u}_1u_2 \\
 -3\lambda \bar{u}_2+i\bar{u}_{2,x} & -iu_1\bar{u}_2 & 3i\lambda^2-i|u_2|^2 \\
 \end{array} \right).
 \label{V}
\end{eqnarray}
\label{UV}
\end{subequations}
Here and in what follows the bar indicates complex conjugate.

\subsection{Equal-time and equal-space Poisson brackets and the associated Hamiltonian forms}

For the NLS equation, it was found in \cite{CK2015,ACDK2016} that there exist two dual Hamiltonian representations:
the first one is defined by the usual equal-time Poisson brackets; while the second one is defined by the new equal-space Poisson brackets. This result was generalised to the vector NLS equation in \cite{zhou2017}.
The Manakov equation (\ref{CNLS}) is corresponding to the vector NLS equation with $N=2$. Thus the new equal-space
Hamiltonian representation for (\ref{CNLS}) was implicit in \cite{zhou2017}.
For self-containedness, in this subsection we revisit the essential results regarding this aspect (for more details we refer the interested reader to \cite{CK2015,ACDK2016,zhou2017}).

It is easy to check that the coupled NLS equation (\ref{CNLS}) can be derived from a variational principle based on the following Lagrangian density
\begin{eqnarray}
\mathcal{L}(u)=\frac{i}{2}\left(\bar{u}_1u_{1,t}-u_1\bar{u}_{1,t}+\bar{u}_2u_{2,t}-u_2\bar{u}_{2,t}\right)
-\left(|u_{1,x}|^2+|u_{2,x}|^2\right)+\left( |u_1|^2+|u_2|^2\right)^2.
\label{CNLSLD}
\end{eqnarray}
We consider the usual equal-time Poisson brackets \cite{Faddeev2007}
\begin{eqnarray}
\begin{split}
\left\{u_j(x,t),u_k(y,t)\right\}_S=\left\{\bar{u}_j(x,t),\bar{u}_k(y,t)\right\}_S=0,
\\
\left\{u_j(x,t),\bar{u}_k(y,t)\right\}_S=-i\delta_{jk}\delta(x-y), ~j,k=1,2,
\end{split}
\label{PBS}
\end{eqnarray}
where $\delta_{jk}$ is the Kronecker $\delta$-function, and $\delta(x-y)$ is the Dirac $\delta$-function.
Based on this equal-time Poisson brackets, the coupled NLS equation can be written as
\begin{eqnarray}
u_{j,t}=\left\{u_j,H_S\right\}_S,~~j,k=1,2,
\label{HFS}
\end{eqnarray}
where $H_S=\int \mathcal{H}_S dx$ with the Hamiltonian density
\begin{eqnarray}
\mathcal{H}_S=\frac{i}{2}\left(\bar{u}_1u_{1,t}-u_1\bar{u}_{1,t}+\bar{u}_2u_{2,t}-u_2\bar{u}_{2,t}\right)-\mathcal{L}(u)
=|u_{1,x}|^2+|u_{2,x}|^2-\left(|u_{1}|^2+|u_{2}|^2\right)^2.
\label{HS}
\end{eqnarray}
The above presentation is the usual space Hamiltonian formulation of the coupled NLS equation.
Let us present a dual version, that is the time Hamiltonian formulation.
The momentum variables corresponding to $u_j$, $\bar{u}_j$ are defined by
\begin{eqnarray}
\frac{\partial \mathcal{L}}{\partial u_{j,x}}=-\bar{u}_{j,x},~~\frac{\partial \mathcal{L}}{\partial \bar{u}_{j,x}}=-u_{j,x}, ~~j=1,2.
\end{eqnarray}
We introduce the new equal-space Poisson brackets for the coupled NLS equation \cite{zhou2017}
\begin{eqnarray}
\begin{split}
\left\{u_j(x,t),\bar{u}_{k,x}(x,\tau)\right\}_T&=\delta_{jk}\delta(t-\tau),
\\
\left\{u_j(x,t),u_{k}(x,\tau)\right\}_T&=\left\{u_j(x,t),\bar{u}_{k}(x,\tau)\right\}_T
=\left\{u_j(x,t),u_{k,x}(x,\tau)\right\}_T=0,
\\
\left\{u_{j,x}(x,t),u_{k,x}(x,\tau)\right\}_T&=\left\{\bar{u}_{j,x}(x,t),\bar{u}_{k,x}(x,\tau)\right\}_T=0,
~~j,k=1,2.
\end{split}
\label{PBT}
\end{eqnarray}
By using these brackets, one can check directly that the coupled NLS equation can be written in the time Hamiltonian form
\begin{eqnarray}
u_{j,x}=\left\{u_{j},H_T\right\}_T,~~u_{j,xx}=\left\{u_{j,x},H_T\right\}_T,~~j,k=1,2,
\label{HFS}
\end{eqnarray}
where $H_T=\int \mathcal{H}_T dt$  with the Hamiltonian density
\begin{eqnarray}
\begin{split}
\mathcal{H}_T&=2\left(|u_{1,x}|^2+|u_{2,x}|^2\right)+\mathcal{L}(u)
\\&=\frac{i}{2}\left(\bar{u}_1u_{1,t}-u_1\bar{u}_{1,t}+\bar{u}_2u_{2,t}-u_2\bar{u}_{2,t}\right)
+|u_{1,x}|^2+|u_{2,x}|^2+\left(|u_{1}|^2+|u_{2}|^2\right)^2.
\end{split}
\label{HT}
\end{eqnarray}

After straightforward calculations using the above two Poisson brackets,
one can deduce that the matrices $U(x, t, \lambda)$ and $V (x, t, \lambda)$
satisfy the same $r$-matrix relation, that is
\begin{eqnarray}
\left\{U_1(x,t,\lambda),U_2(y,t,\mu)\right\}_S=\left[r(\lambda-\mu),U_1(x,t,\lambda)+U_2(y,t,\mu)\right]\delta(x-y),
\label{Urmrelation}
\\
\left\{V_1(x,t,\lambda),V_2(x,\tau,\mu)\right\}_T=\left[r(\lambda-\mu),V_1(x,t,\lambda)+V_2(x,\tau,\mu)\right]\delta(t-\tau),
\label{Vrmrelation}
\end{eqnarray}
where $U_1(x,t,\lambda)=U(x,t,\lambda) \otimes I_3$, $U_2(y,t,\mu)=I_3\otimes U(y,t,\mu)$, $V_1(x,t,\lambda)=V(x,t,\lambda) \otimes I_3$, $V_2(x,\tau,\mu)=I_3\otimes V(x,\tau,\mu)$,
$I_n$ being the $n\times n$ identity matrix, and the classical $r$-matrix is given by
\begin{eqnarray}
r(\lambda)=\frac{1}{3\lambda}\mathcal{P},~~\mathcal{P}=\sum_{j,k=1}^{3}e_{jk}\otimes e_{kj},
 \label{rm}
\end{eqnarray}
and $e_{jk}$ denotes $3\times 3$ matrix having $1$ in the $(j,k)$-th position and zeros elsewhere.
Note that the permutation matrix $\mathcal{P}$ satisfies
\begin{eqnarray}
\mathcal{P}\left(A\otimes B\right)=\left(B\otimes A\right)\mathcal{P}
\label{Prel}
\end{eqnarray}
where $A$ and $B$ are $3\times 3$ matrices.

\begin{remark} It is worth to point out that, following the original derivation in the scalar case \cite{CK2015,ACDK2016},
there should have a relative minus sign between the $r$-matrix formulations (\ref{Urmrelation}) and (\ref{Vrmrelation}). This sign has an important interpretation from the point of view of the covariant $r$-matrix structure \cite{Caudrelier2020}.
However this property is not essential for us to study the integrability of the type I defect system.
For convenience, here we have dispensed with this sign by exchanging the positions of $u_j$ and $\bar{u}_{j,x}$ in equal-space Poisson brackets (\ref{PBT}).
\end{remark}

\subsection{Integrability of the coupled NLS equation with respect to the two Poisson brackets}

Based on the equal-time Poisson brackets (\ref{PBS}) and the Poisson bracket matrix (\ref{Urmrelation}),
one can implement the standard $r$-matrix approach \cite{Faddeev2007} to study the Liouville integrability of the coupled NLS equation. The analysis goes as follows. For fixing the idea, we focus on the vanishing boundary conditions as $x$ going to infinity (the discussion for the periodic boundary conditions as $x$ posed on a finite interval is essentially the same).
We introduce the transition matrix \cite{Faddeev2007}
\begin{eqnarray}
M_S(x,y,\lambda)=\overset{\curvearrowleft}\exp \int_{y}^{x} U(\xi,t,\lambda)d\xi.
\label{MS}
\end{eqnarray}
The relation (\ref{Urmrelation}) implies the following Poisson brackets between the entries of the transition matrix  (see for example \cite{Faddeev2007,DFR})
\begin{eqnarray}
\left\{M_{S1}(x,y,\lambda),M_{S2}(x,y,\mu)\right\}_S=\left[r(\lambda-\mu),M_{S1}(x,y,\lambda)M_{S2}(x,y,\mu)\right],
\label{rmrelation}
\end{eqnarray}
where $M_{S1}(x,y,\lambda)=M_S(x,y,\lambda)\otimes I_3$, $M_{S2}(x,y,\mu)=I_3\otimes M_S(x,y,\mu)$.
For the problem with vanishing boundary conditions at infinity, this relation implies the Liouville integrability of the model in the sense that
there exists an infinite set of Poisson commuting integrals of motion. Indeed, (\ref{rmrelation}) implies that $\tr(M_S(\lambda))$ commutes for different values of the spectral parameter:
\begin{eqnarray}
\left\{\tr(M_S(\lambda)),\tr(M_S(\mu))\right\}_S=0,
\end{eqnarray}
where $M_S(\lambda)=M_S(\infty,-\infty,\lambda)$.
Thus $\tr(M_S(\lambda))$ provides a generating function of integrals of motion which commutes each other with respect to the equal-time Poisson brackets (\ref{PBS}).
The explicit forms for these integrals of motion can be derived by studying the large $\lambda$ asymptotic expansion of $\tr(M_S(\lambda))$. We refer the reader to \cite{Faddeev2007,DFR} for details regarding this subject.
Here, for our purposes, it is more convenient for us to derive the same result directly from the Lax pair formulation. We denote
\begin{eqnarray}
\Gamma_1=\frac{\phi_2}{\phi_1},~~\Gamma_2=\frac{\phi_3}{\phi_1}.
\end{eqnarray}
From (\ref{LPxt}), we find the following space-part Riccati equations
\begin{subequations}
\begin{eqnarray}
\Gamma_{1,x}=3i\lambda \Gamma_1-u_1\left(\Gamma_1\right)^2-u_2\Gamma_1\Gamma_2-\bar{u}_1,
\\
\Gamma_{2,x}=3i\lambda \Gamma_2-u_2\left(\Gamma_2\right)^2-u_1\Gamma_1\Gamma_2-\bar{u}_2,
\end{eqnarray}
\label{ricx}
\end{subequations}
and the following time-part Riccati equations
\begin{subequations}
\begin{eqnarray}
\Gamma_{1,t}=V_{21}+\left(V_{22}-V_{11}\right)\Gamma_1+V_{23}\Gamma_2-V_{12}\left(\Gamma_1\right)^2- V_{13}\Gamma_1\Gamma_2,
\\
\Gamma_{2,t}=V_{31}+\left(V_{33}-V_{11}\right)\Gamma_2+V_{32}\Gamma_1-V_{12}\Gamma_1\Gamma_2- V_{13}\left(\Gamma_2\right)^2,
\end{eqnarray}
\label{rict}
\end{subequations}
where $V_{jk}$, $1\leq j,k\leq3$, denote the $jk$-entries of the matrix $V(x,t,\lambda)$.
From the compatibility condition $\left(\ln\phi_1\right)_{xt}=\left(\ln\phi_1\right)_{tx}$, we find the following conservation equation
\begin{eqnarray}
\left(u_1\Gamma_1+u_2\Gamma_2\right)_t=
\left(V_{11}+V_{12}\Gamma_1+V_{13}\Gamma_2\right)_x.
\label{CL}
\end{eqnarray}
Using the vanishing boundary condition, equation (\ref{CL}) implies that the function $u_1\Gamma_1+u_2\Gamma_2$ provides a generating function of the conservation densities.
By substituting the expansion
\begin{eqnarray}
\Gamma_1=\sum_{n=1}^{\infty}\Gamma_1^{(n)}(3i\lambda)^{-n}, ~~\Gamma_2=\sum_{n=1}^{\infty}\Gamma_2^{(n)}(3i\lambda)^{-n}
\label{Gamexpansion}
\end{eqnarray}
into (\ref{ricx}) and by equating the coefficients of powers of $\lambda$,
we find explicit forms of $\Gamma_1^{(n)}$ and $\Gamma_2^{(n)}$ as follows
\begin{eqnarray}
\begin{split}
\Gamma_{1}^{(1)}&=\bar{u}_1,~~\Gamma_{2}^{(1)}=\bar{u}_2, ~~\Gamma_{1}^{(2)}=\bar{u}_{1,x},~~\Gamma_{2}^{(2)}=\bar{u}_{2,x}, \\
\Gamma_{1}^{(3)}&=\bar{u}_{1,xx}+\bar{u}_1\left(|u_{1}|^2+|u_{2}|^2\right),
~~\Gamma_{2}^{(3)}=\bar{u}_{2,xx}+\bar{u}_2\left(|u_{1}|^2+|u_{2}|^2\right),\\
\Gamma_1^{(n+1)}&=\left(\Gamma_{1}^{(n)}\right)_x+u_1\sum_{j=1}^{n-1}\Gamma_1^{(j)}\Gamma_1^{(n-j)}
+u_2\sum_{j=1}^{n-1}\Gamma_1^{(j)}\Gamma_2^{(n-j)},\quad n\geq 2,\\
\Gamma_2^{(n+1)}&=\left(\Gamma_{2}^{(n)}\right)_x+u_2\sum_{j=1}^{n-1}\Gamma_2^{(j)}\Gamma_2^{(n-j)}
+u_1\sum_{j=1}^{n-1}\Gamma_1^{(j)}\Gamma_2^{(n-j)},\quad n\geq 2.
\end{split}
\label{wj}
\end{eqnarray}
Thus the integrals of motion are given by
\begin{eqnarray}
I_n=\int_{-\infty}^{\infty}\left(u_1\Gamma^{(n)}_1+u_2\Gamma^{(n)}_2\right)dx, ~~n\geq 1.
\label{In}
\end{eqnarray}
The Hamiltonian $H_S$ defined by (\ref{HS}) is generated by $I_3$ after an integration by parts.

We now show how to study the integrability of the coupled NLS equation from
the point of view of the new equal-space Poisson brackets (\ref{PBT}).
The analysis follows the same line as the analogous problem in the scalar case \cite{CK2015}.
The starting point is the Poisson bracket matrix (\ref{Vrmrelation}).
Relation (\ref{Vrmrelation}) implies that the transition matrix
\begin{eqnarray}
M_T(t,\tau,\lambda)=\overset{\curvearrowleft}\exp \int_{\tau}^{t} V(x,\eta,\lambda)d\eta
\label{MT}
\end{eqnarray}
satisfies the following $r$-matrix relation
\begin{eqnarray}
\left\{M_{T1}(t,\tau,\lambda),M_{T2}(t,\tau,\mu)\right\}_T=\left[r(\lambda-\mu),M_{T1}(t,\tau,\lambda)M_{T2}(t,\tau,\mu)\right].
\label{rmrelationT}
\end{eqnarray}
This implies that $\tr(M_T(\lambda))$, $M_T(\lambda)=M_T(\infty,-\infty,\lambda)$, generates the infinite set of conserved quantities (conserved with respect to $x$) which commutes each other with respect to the equal-space Poisson brackets (\ref{PBT}).
Following the same reasoning as in the case of equal-time Poisson brackets, we use the conservation equation (\ref{CL}) to extract these conserved quantities. More precisely, equation (\ref{CL}) implies that a generating function of the conservation densities in space  is given by the function $V_{11}+V_{12}\Gamma_1+V_{13}\Gamma_2$.
By substituting the expansion
\begin{eqnarray}
\Gamma_1=\sum_{n=1}^{\infty}\gamma_1^{(n)}(3i\lambda)^{-n}, ~~\Gamma_2=\sum_{n=1}^{\infty}\gamma_2^{(n)}(3i\lambda)^{-n}
\label{Gamexpansiont}
\end{eqnarray}
into the time Ricatti equation (\ref{rict}), we find
\begin{eqnarray}
\begin{split}
\gamma_{1}^{(1)}=&\bar{u}_1,~~\gamma_{2}^{(1)}=\bar{u}_2, ~~\gamma_{1}^{(2)}=\bar{u}_{1,x},~~\gamma_{2}^{(2)}=\bar{u}_{2,x},\\ \gamma_{1}^{(3)}=&i\bar{u}_{1,t}-\left(|u_1|^2+|u_2|^2\right)\bar{u}_1,
~~\gamma_{2}^{(3)}=i\bar{u}_{2,t}-\left(|u_1|^2+|u_2|^2\right)\bar{u}_2,
\\
\gamma_1^{(n+2)}=&i\gamma_{1,t}^{(n)}-\left(2|u_1|^2+|u_2|^2\right)\gamma_1^{(n)}
-\bar{u}_1u_2\gamma_2^{(n)}+u_1\sum_{j+k=n+1}\gamma_1^{(j)}\gamma_1^{(k)}
\\&+u_2\sum_{j+k=n+1}\gamma_1^{(j)}\gamma_2^{(k)}
-u_{1,x}\sum_{j+k=n}\gamma_1^{(j)}\gamma_1^{(k)}
-u_{2,x}\sum_{j+k=n}\gamma_1^{(j)}\gamma_2^{(k)},~~n\geq 1,
\\
\gamma_2^{(n+2)}=&i\gamma_{2,t}^{(n)}-\left(|u_1|^2+2|u_2|^2\right)\gamma_2^{(n)}
-u_1\bar{u}_2\gamma_1^{(n)}+u_1\sum_{j+k=n+1}\gamma_1^{(j)}\gamma_2^{(k)}
\\&+u_2\sum_{j+k=n+1}\gamma_2^{(j)}\gamma_2^{(k)}
-u_{1,x}\sum_{j+k=n}\gamma_1^{(j)}\gamma_2^{(k)}
-u_{2,x}\sum_{j+k=n}\gamma_2^{(j)}\gamma_2^{(k)},~~n\geq 1.
\end{split}
\label{wjt}
\end{eqnarray}
We write
\begin{eqnarray*}
V_{11}+V_{12}\Gamma_1+V_{13}\Gamma_2=-6i\lambda^2+\sum_{n=1}^{\infty}\mathcal{K}_n(3i\lambda)^{-n},
\end{eqnarray*}
then the corresponding integrals are
\begin{eqnarray}
K_n=\int_{-\infty}^{\infty}\mathcal{K}_ndt,
~~\mathcal{K}_n=i\left(u_{1,x}\gamma_1^{(n)}+u_{2,x}\gamma_2^{(n)}-u_1\gamma_1^{(n+1)}-u_2\gamma_2^{(n+1)}\right).
\end{eqnarray}
These integrals are in involution with respect to the equal-space Poisson brackets (\ref{PBT}).
In particular, we have
\begin{eqnarray}
\mathcal{H}_T=-\frac{i}{2}\left(\mathcal{K}_2-\bar{\mathcal{K}}_2\right).
\end{eqnarray}
Thus, we recover the Hamiltonian $H_T$ given by (\ref{HT}).

\begin{remark} The analyses in this section show that the usual equal-space and the new equal-time Poisson brackets provide completely equivalent Hamiltonian descriptions for the coupled NLS equation without a defect.
However, the advantages of using the new Poisson structure lie in that it enables us to interpret the defect condition described by a frozen BT simply as a canonical transformation, and it enables us to implement the classical $r$-matrix method to prove Liouville integrability of the defect coupled NLS equation;
this fact will become clearly in the next two sections.
\end{remark}

\section{A BT for the coupled NLS equation and its canonical property}

\subsection{A BT for the coupled NLS equation}

As we have mentioned above, we will use the idea of BTs as defect conditions.
So, we first need to derive a BT that is suitable for describing the type I defect condition for the coupled NLS equation. This will be fulfilled in this subsection.

We introduce another copy of the auxiliary problems for $\tilde{\phi}$ with Lax pair
$\tilde{U}$, $\tilde{V}$ defined as in (\ref{UV}) with the new fields $\tilde{u}_1$, $\tilde{u}_2$, replacing $u_1$, $u_2$.
We assume that the two systems are related by the gauge transformation,
\begin{eqnarray}
\phi(x,t,\lambda)=B(x,t,\lambda)\tilde{\phi}(x,t,\lambda),
\label{BT}
\end{eqnarray}
where the matrix $B(x,t,\lambda)$ satisfies
\begin{subequations}
\begin{eqnarray}
B_x(x,t,\lambda)=U(x,t,\lambda)B(x,t,\lambda)-B(x,t,\lambda)\tilde{U}(x,t,\lambda),
\label{BT1a}
\\
B_t(x,t,\lambda)=V(x,t,\lambda)B(x,t,\lambda)-B(x,t,\lambda)\tilde{V}(x,t,\lambda).
\label{BT1b}
\end{eqnarray}
\label{BT1}
\end{subequations}
In order to describe a type I defect system, we need to find a matrix $B$ such that it dependents on the spectral parameter $\lambda$ and the fields $u_j$ and $\tilde{u}_j$, $j=1,2$, but not on extra degrees of freedom. We will show below such a BT matrix can be derived by using the well-known Darboux transformation (DT) for the coupled NLS equation \cite{Wright2003,Gu}. Indeed, the DT matrix for the coupled NLS equation is given by \cite{Wright2003,Gu}
\begin{eqnarray}
 T=\left(\lambda-\bar{\xi}_{1}\right) I_3+\frac{\bar{\xi}_{1}-\xi_1}{|f_{1}|^2+|f_{2}|^2+|f_{3}|^2}\left( \begin{array}{ccc} |f_{1}|^2 & f_1\bar{f}_2 & f_1\bar{f}_3\\
 f_2\bar{f}_1 &  |f_{2}|^2 & f_2\bar{f}_3\\
f_3\bar{f}_1 & f_3\bar{f}_2  & |f_{3}|^2
 \end{array} \right)\equiv \left(\lambda-\bar{\xi}_{1}\right) I_3+T^{(0)},
\label{DTM}
\end{eqnarray}
where $(f_{1},f_{2},f_{3})^T$ is a special solution of the linear auxiliary systems (\ref{LPxt}) with $\lambda=\xi_1$ and with $\tilde{u}_j$, $j=1,2$. The transformation between the fields is
\begin{subequations}
\begin{eqnarray}
u_{1}-\tilde{u}_{1}=6\im\left(\xi_1\right)\frac{f_{1}\bar{f}_{2}}{|f_{1}|^2+|f_{2}|^2+|f_{3}|^2},
\label{DTa}
\\
u_{2}-\tilde{u}_{2}=6\im\left(\xi_1\right)\frac{f_{1}\bar{f}_{3}}{|f_{1}|^2+|f_{2}|^2+|f_{3}|^2}.
\label{DTb}
\end{eqnarray}
\label{DT}
\end{subequations}
By using (\ref{DT}), we are able to express the entries of $T^{(0)}$ in terms of the fields $u_j$ and $\tilde{u}_j$, $j=1,2$. More precisely, we obtain
\begin{eqnarray}
\begin{split}
T^{(0)}_{12}=\frac{i}{3}(\tilde{u}_1-u_1),~~T^{(0)}_{13}=\frac{i}{3}(\tilde{u}_2-u_2),
~~T^{(0)}_{21}=-\bar{T}^{(0)}_{12},~~T^{(0)}_{31}=-\bar{T}^{(0)}_{13},\\
\left(T^{(0)}_{11}+i\im(\xi_1)\right)^2=-\frac{1}{9}\left(9\left(\im(\xi_1)\right)^2-|u_1-\tilde{u}_1|^2-|u_2-\tilde{u}_2|^2\right),\\
T^{(0)}_{22}=\frac{T^{(0)}_{12}T^{(0)}_{21}}{T^{(0)}_{11}},~~T^{(0)}_{33}=\frac{T^{(0)}_{13}T^{(0)}_{31}}{T^{(0)}_{11}},
~~T^{(0)}_{23}=\frac{T^{(0)}_{21}T^{(0)}_{13}}{T^{(0)}_{11}},~~T^{(0)}_{32}=-\bar{T}^{(0)}_{23}.
\end{split}
\label{T0}
\end{eqnarray}
For fixing the idea, we will take
$$T^{(0)}_{11}+i\im(\xi_1)=\frac{i}{3}\Omega, ~~\Omega= \sqrt{9\left(\im(\xi_1)\right)^2-|u_1-\tilde{u}_1|^2-|u_2-\tilde{u}_2|^2},$$
in what follows.
It is convenient for us to multiply the DT matrix (\ref{DTM}) with the factor $\lambda^{-1}$ and to take the notation $\xi_1=-(a+ib)$, $a,b\in \mathbb{R}$. Combining the above discussion, we obtain the following BT matrix
\begin{eqnarray}
B(x,t,\lambda)= I_3+\lambda^{-1}B^{(0)}(x,t),
 \label{cnlsdm}
\end{eqnarray}
where the entries of the $3\times 3$ matrix $B^{(0)}(x,t)$ are given by
\begin{eqnarray}
\begin{split}
B^{(0)}_{11}&=a+\frac{i}{3}\Omega,~~\Omega= \sqrt{9b^2-|u_1-\tilde{u}_1|^2-|u_2-\tilde{u}_2|^2},\\
B^{(0)}_{12}&=\frac{i}{3}(\tilde{u}_1-u_1),~~B^{(0)}_{13}=\frac{i}{3}(\tilde{u}_2-u_2),~~B^{(0)}_{21}=-\bar{B}^{(0)}_{12},\\
B^{(0)}_{22}&=a-ib-\frac{|B^{(0)}_{12}|^2}{\frac{i}{3}\Omega+ib},~~B^{(0)}_{23}=\frac{B^{(0)}_{21}B^{(0)}_{13}}{\frac{i}{3}\Omega+ib},\\
B^{(0)}_{31}&=-\bar{B}^{(0)}_{13},~~B^{(0)}_{32}=-\bar{B}^{(0)}_{23},~~B^{(0)}_{33}=a-ib-\frac{|B^{(0)}_{13}|^2}{\frac{i}{3}\Omega+ib},
\end{split}
\label{BTM}
\end{eqnarray}
with $a$ and $b$ being real parameters.

Inserting (\ref{cnlsdm}) into equations (\ref{BT1}), we obtain the following transformation between the fields $\left(u_1(x,t), u_2(x,t)\right)$ and $\left(\tilde{u}_1(x,t),\tilde{u}_2(x,t)\right)$:
\begin{eqnarray}
\begin{split}
\left(u_{1}-\tilde{u}_{1}\right)_x=&\tilde{u}_1\left(\Omega-3ia\right)+3iu_1B^{(0)}_{22}+3iu_2B^{(0)}_{32},
\\
\left(u_{2}-\tilde{u}_{2}\right)_x=&\tilde{u}_2\left(\Omega-3ia\right)+3iu_1B^{(0)}_{23}+3iu_2B^{(0)}_{33},
\\
\left(u_{1}-\tilde{u}_{1}\right)_t=&i(u_1-\tilde{u}_1)\Delta
+i\left(\tilde{u}_1u_2-u_1\tilde{u}_2\right)\bar{\tilde{u}}_2+i\tilde{u}_{1,x}\left(\Omega-3ia\right)-3u_{1,x}B^{(0)}_{22}-3u_{2,x}B^{(0)}_{32},
\\
\left(u_{2}-\tilde{u}_{2}\right)_t=&i(u_2-\tilde{u}_2)\Delta
+i(u_1\tilde{u}_2-\tilde{u}_1u_2)\bar{\tilde{u}}_1+i\tilde{u}_{2,x}\left(\Omega-3ia\right)-3u_{1,x}B^{(0)}_{23}-3u_{2,x}B^{(0)}_{33},
\end{split}
\label{BTCNLS}
\end{eqnarray}
where $\Omega$, $B^{(0)}_{jk}$ are defined by (\ref{BTM}), and $\Delta=|u_1|^2+|u_2|^2+|\tilde{u}_1|^2+|\tilde{u}_2|^2$. This is the desired BT for describing our type I defect system for the coupled NLS equation.
To the best of our knowledge, the BT in the form of (\ref{BTCNLS}) and the associated BT matrix (\ref{cnlsdm}) have not yet been presented explicitly in the literature, despite the DTs for the coupled NLS have been well investigated.
We will refer to the first two equations and the second two equations in (\ref{BTCNLS}) as the space-parts and the time-parts of the BT, respectively.

\subsection{Canonical property of the BT}

Canonical properties of BTs with respect to the usual equal-time Poisson brackets were first established in \cite{KW1976,K1977}.
These results were generalized in \cite{CK2015} to the NLS equation with respect to the new equal-space Poisson brackets, in order to study the integrability of the defect system.
Here we follow the method of \cite{CK2015} in the context of NLS equation and generalise it to our coupled case:
we will show that the BT (\ref{BTCNLS}) for the coupled NLS equation is also a canonical transformation with respect to both the usual equal-time Poisson brackets (\ref{PBS}) and the new equal-space Poisson brackets (\ref{PBT}).

We first give the following important result.
\begin{lemma}
For two systems $(u_1,u_2)$ and $(\tilde{u}_1,\tilde{u}_2)$ connected by the BT,
the one forms
\begin{eqnarray}
\begin{split}
\omega&=\left(u_1\Gamma_1+u_2\Gamma_2\right)dx
+\left(V_{11}+V_{12}\Gamma_1+V_{13}\Gamma_2\right)dt,
\\
\tilde{\omega}&=\left(\tilde{u}_1\tilde{\Gamma}_1+\tilde{u}_2\tilde{\Gamma}_2\right)dx
+\left(\tilde{V}_{11}+\tilde{V}_{12}\tilde{\Gamma}_1+\tilde{V}_{13}\tilde{\Gamma}_2\right)dt
\end{split}
\end{eqnarray}
differ only by an exact form. More precisely,
\begin{eqnarray}
\omega-\tilde{\omega}=d\left(\ln\left(B_{11}+B_{12}\tilde{\Gamma}_1+B_{13}\tilde{\Gamma}_2\right)\right),
\label{omgomgt}
\end{eqnarray}
where $B_{jk}$, $1\leq j,k\leq3$, denote the $jk$-entries of the BT matrix $B(x,t,\lambda)$.
\end{lemma}
{\bf Proof}
The formula (\ref{omgomgt}) is equivalent to
\begin{eqnarray}
u_1\Gamma_1+u_2\Gamma_2
-\left(\tilde{u}_1\tilde{\Gamma}_1+\tilde{u}_2\tilde{\Gamma}_2\right)=
\partial_x\ln\left(B_{11}+B_{12}\tilde{\Gamma}_1+B_{13}\tilde{\Gamma}_2\right),
\label{DU}
\\
V_{11}+V_{12}\Gamma_1+V_{13}\Gamma_2
-\left(\tilde{V}_{11}+\tilde{V}_{12}\tilde{\Gamma}_1+\tilde{V}_{13}\tilde{\Gamma}_2\right)
=\partial_t\ln\left(B_{11}+B_{12}\tilde{\Gamma}_1+B_{13}\tilde{\Gamma}_2\right).
\label{DV}
\end{eqnarray}
We first prove the formula (\ref{DU}).
From (\ref{BT}), we have
\begin{subequations}
\begin{eqnarray}
\Gamma_1= \frac{B_{21}+B_{22}\tilde{\Gamma}_1+B_{23}\tilde{\Gamma}_2}
{B_{11}+B_{12}\tilde{\Gamma}_1+B_{13}\tilde{\Gamma}_2},
\\
\Gamma_2=\frac{B_{31}+B_{32}\tilde{\Gamma}_1+B_{33}\tilde{\Gamma}_2}
{B_{11}+B_{12}\tilde{\Gamma}_1+B_{13}\tilde{\Gamma}_2}.
\end{eqnarray}
\label{gtildeg}
\end{subequations}
By using (\ref{gtildeg}) to eliminate $\Gamma_1$ and $\Gamma_2$,
the left hand side of (\ref{DU}) can be written as
\begin{eqnarray}
\begin{split}
\frac{1}{F_1}\left(u_1F_2+u_2F_3-\left(\tilde{u}_{1}\tilde{\Gamma}_1+\tilde{u}_{2}\tilde{\Gamma}_2\right)F_1\right),
\end{split}
\label{der11x}
\end{eqnarray}
where $F_j=B_{j1}+B_{j2}\tilde{\Gamma}_1+B_{j3}\tilde{\Gamma}_2$, $j=1,2,3$.
The right hand side of (\ref{DU}) reads
\begin{eqnarray}
\frac{\left(B_{11}+B_{12}\tilde{\Gamma}_1+B_{13}\tilde{\Gamma}_2\right)_x}{B_{11}+B_{12}\tilde{\Gamma}_1+B_{13}\tilde{\Gamma}_2}.
\label{der2x}
\end{eqnarray}
We note that (\ref{BT1a}) implies
\begin{subequations}
\begin{eqnarray}
\left(B_{11}\right)_x=u_1B_{21}+u_2B_{31}+\tilde{\bar{u}}_1B_{12}+\tilde{\bar{u}}_2B_{13},
\\
\left(B_{12}\right)_x=-3i\lambda B_{12}+u_1B_{22}+u_2B_{32}-\tilde{u}_1B_{11},
\\
\left(B_{13}\right)_x=-3i\lambda B_{13}+u_1B_{23}+u_2B_{33}-\tilde{u}_2B_{11}.
\end{eqnarray}
\label{BT11x}
\end{subequations}
Using (\ref{BT11x}) and the space-part Riccati equations
\begin{subequations}
\begin{eqnarray}
\tilde{\Gamma}_{1,x}=3i\lambda\tilde{\Gamma}_1-\tilde{u}_{1}\left(\tilde{\Gamma}_1\right)^2 -\tilde{u}_{2}\tilde{\Gamma}_1\tilde{\Gamma}_2- \tilde{\bar{u}}_{1},
\\
\tilde{\Gamma}_{2,x}=3i\lambda\tilde{\Gamma}_2-\tilde{u}_{2}\left(\tilde{\Gamma}_2\right)^2 -\tilde{u}_{1}\tilde{\Gamma}_1\tilde{\Gamma}_2- \tilde{\bar{u}}_{2},
\end{eqnarray}
\label{ricttildex}
\end{subequations}
one may check directly that (\ref{der2x}) is equal to (\ref{der11x}).
Thus the formula (\ref{DU}) holds.
The proof for the formula (\ref{DV}) can be performed via a similar manner.
More precisely, by using (\ref{gtildeg}), the left hand side of (\ref{DV}) can be written as
\begin{eqnarray}
\begin{split}
\frac{1}{F_1}\left(\left(V_{11}-\tilde{V}_{11}-\tilde{V}_{12}\tilde{\Gamma}_1-\tilde{V}_{13}\tilde{\Gamma}_2\right)F_1
 +V_{12}F_2+V_{13}F_3\right).
\end{split}
\label{der11}
\end{eqnarray}
The right hand side of (\ref{DV}) is equal to
\begin{eqnarray}
\frac{\left(B_{11}+B_{12}\tilde{\Gamma}_1+B_{13}\tilde{\Gamma}_2\right)_t}{B_{11}+B_{12}\tilde{\Gamma}_1+B_{13}\tilde{\Gamma}_2}.
\label{der2}
\end{eqnarray}
We note that (\ref{BT1b}) implies
\begin{subequations}
\begin{eqnarray}
\left(B_{11}\right)_t=V_{11}B_{11}+V_{12}B_{21}+V_{13}B_{31}-B_{11}\tilde{V}_{11}-B_{12}\tilde{V}_{21}-B_{13}\tilde{V}_{31},
\\
\left(B_{12}\right)_t=V_{11}B_{12}+V_{12}B_{22}+V_{13}B_{32}-B_{11}\tilde{V}_{12}-B_{12}\tilde{V}_{22}-B_{13}\tilde{V}_{32},
\\
\left(B_{13}\right)_t=V_{11}B_{13}+V_{12}B_{23}+V_{13}B_{33}-B_{11}\tilde{V}_{13}-B_{12}\tilde{V}_{23}-B_{13}\tilde{V}_{33}.
\end{eqnarray}
\label{BT11}
\end{subequations}
Using (\ref{BT11}) and the time-part Riccati equations
\begin{subequations}
\begin{eqnarray}
\tilde{\Gamma}_{1,t}=\tilde{V}_{21}+\left(\tilde{V}_{22}-\tilde{V}_{11}\right)\tilde{\Gamma}_1
+\tilde{V}_{23}\tilde{\Gamma}_2-\tilde{V}_{12}\left(\tilde{\Gamma}_1\right)^2
- \tilde{V}_{13}\tilde{\Gamma}_1\tilde{\Gamma}_2,
\\
\tilde{\Gamma}_{2,t}=\tilde{V}_{31}+\left(\tilde{V}_{33}-\tilde{V}_{11}\right)\tilde{\Gamma}_2
+\tilde{V}_{32}\tilde{\Gamma}_1-\tilde{V}_{12}\tilde{\Gamma}_1\tilde{\Gamma}_2- \tilde{V}_{13}\left(\tilde{\Gamma}_2\right)^2,
\end{eqnarray}
\label{ricttilde}
\end{subequations}
one may check directly that (\ref{der2}) is equal to (\ref{der11}). This completes the proof.
\QEDB

Formula (\ref{DU}) implies that the conserved densities (in space) $\mathcal{I}_n$ and $\tilde{\mathcal{I}}_n$ of the two systems differ only by a total space derivative of some functional. That is
\begin{eqnarray}
\mathcal{I}_n=\tilde{\mathcal{I}}_n+\partial_x \mathcal{F}_n, ~~n\geq 1,
\label{InD}
\end{eqnarray}
where $\mathcal{F}_n$ is determined by the following expansion
\begin{eqnarray}
\ln\left(B_{11}+B_{12}\tilde{\Gamma}_1+B_{13}\tilde{\Gamma}_2\right)=\sum_{n=1}^{\infty} \frac{\mathcal{F}_n}{(3i\lambda)^n}.
\end{eqnarray}
Analogously, formula (\ref{DV}) implies that the conserved densities (in time) $\mathcal{K}_n$ and $\tilde{\mathcal{K}}_n$  of the two systems differ only by a total time derivative of some functional.
That is
\begin{eqnarray}
\mathcal{K}_n=\tilde{\mathcal{K}}_n+\partial_t \mathcal{F}_n, ~~n\geq 1.
\label{KnD}
\end{eqnarray}
In particular, for the Hamiltonian densities $(\mathcal{H}_S, \tilde{\mathcal{H}}_S)$ and $(\mathcal{H}_T, \tilde{\mathcal{H}}_T)$, we have
\begin{eqnarray}
\begin{split}
\mathcal{H}_S=&\tilde{\mathcal{H}}_S
+\partial_x\left(\frac{\Omega^3}{3}+\Omega\left(|\tilde{u}_1|^2+|\tilde{u}_2|^2-u_1\bar{\tilde{u}}_1
-u_2\bar{\tilde{u}}_2\right)+u_1\left(\bar{u}_1-\bar{\tilde{u}}_1\right)_x
+u_2\left(\bar{u}_2-\bar{\tilde{u}}_2\right)_x\right)
\\&+\partial_x\left(3ai\left(|u_1|^2+|u_2|^2-\bar{u}_1\tilde{u}_1-\bar{u}_2\tilde{u}_2\right)-9a^2\Omega\right),
\end{split}
\label{HSD}
\end{eqnarray}
and
\begin{eqnarray}
\begin{split}
\mathcal{H}_T=\tilde{\mathcal{H}}_T
+\partial_t\left(\frac{i}{2}\left(\bar{u}_1\tilde{u}_1-u_1\bar{\tilde{u}}_1+\bar{u}_2\tilde{u}_2-u_2\bar{\tilde{u}}_2-6ia\Omega\right)\right).
\end{split}
\label{HTD}
\end{eqnarray}
Equations (\ref{InD}) and (\ref{HSD}) imply that the BT preserves the form of the Hamiltonian as well as that of all local conserved quantities for the coupled NLS equation with the usual equal-time Poisson brackets (\ref{PBS}).
Equations (\ref{KnD}) and (\ref{HTD}) imply that the BT preserves the form of the Hamiltonian as well as that of all local conserved quantities for the coupled NLS equation with the new equal-space Poisson brackets (\ref{PBT}).
Thus our BT (\ref{BTCNLS}) is indeed a canonical transformation with respect to both two Poisson brackets.
We will employ such a canonical property to establish the integrability of the defect system in the next section.

\section{An integrable defect for the coupled NLS equation}

\subsection{The defect system and its Lagrangian description}

We now consider the coupled NLS equation with a type I defect at a specific point in space.
We follow the idea of using BTs as defect conditions \cite{BCZ20041}.
Without loss of generality, we suppose that a defect is located at $x=0$.
\begin{definition}
The coupled NLS equation with a defect located at $x=0$ is described by the following internal boundary problem:
\begin{itemize}
\item $u_1(x,t)$, $u_2(x,t)$ and $\tilde{u}_1(x,t)$, $\tilde{u}_2(x,t)$ satisfy the coupled NLS equation (\ref{CNLS}) in the bulk regions $x>0$ and $x<0$, respectively;
\item at $x=0$, the fields $u_1(x,t)$, $u_2(x,t)$ and $\tilde{u}_1(x,t)$, $\tilde{u}_2(x,t)$ are connected by a condition corresponding to the BT (\ref{BTCNLS}) evaluated at $x=0$. This condition is referred to as defect condition.
\end{itemize}
\end{definition}

We claim that the coupled NLS equation in the presence of the above defined defect also admits a Lagrangian description.
To show this, we introduce the new Lagrangian
\begin{eqnarray}
L=\int_{-\infty}^{0}dx \mathcal{L}(\tilde{u})+\left.D\right|_{x=0}+\int_{0}^{\infty}dx \mathcal{L}(u),
\label{LNLS}
\end{eqnarray}
where $\mathcal{L}(u)$ and $\mathcal{L}(\tilde{u})$ are the Lagrangian densities for the bulk regions $x>0$ and
$x<0$ (see the formula (\ref{CNLSLD})), and $\left.D\right|_{x=0}$ represents a defect contribution at $x=0$.
We will assume that the functional $D$ depends on $u_j$, $\tilde{u}_j$, $u_{j,t}$ and $\tilde{u}_{j,t}$, $j=1,2$, and not on the spatial derivatives.
We consider the complete action
\begin{eqnarray}
\mathcal{A}=\int_{-\infty}^{\infty} dt\left\{\int_{-\infty}^{0}dx \mathcal{L}(\tilde{u})+\left.D\right|_{x=0}+\int_{0}^{\infty}dx \mathcal{L}(u)\right\},
\label{CNLSA}
\end{eqnarray}
and its variations with respect to $\bar{u}_j$ and $\bar{\tilde{u}}_j$, $j=1,2$.
By requiring the variations of (\ref{CNLSA}) with respect to $\bar{u}_j$ and $\bar{\tilde{u}}_j$ to be stationary,
we find
\begin{subequations}
\begin{eqnarray}
\frac{\partial\mathcal{L}(u)}{\partial \bar{u}_j}-\frac{\partial}{\partial x}\left(\frac{\partial\mathcal{L}(u)}{\partial \bar{u}_{j,x}}\right)-\frac{\partial}{\partial t}\left(\frac{\partial\mathcal{L}(u)}{\partial \bar{u}_{j,t}}\right)=0, ~~ x>0,~~j=1,2,
\label{NLSa}
\\
\frac{\partial\mathcal{L}(u)}{\partial \bar{\tilde{u}}_j}-\frac{\partial}{\partial x}\left(\frac{\partial\mathcal{L}(u)}{\partial \bar{\tilde{u}}_{j,x}}\right)-\frac{\partial}{\partial t}\left(\frac{\partial\mathcal{L}(u)}{\partial \bar{\tilde{u}}_{j,t}}\right)=0, ~~ x<0,~~j=1,2,
\label{NLSb}
\\
u_{j,x}+\frac{\partial D}{\partial \bar{u}_j}-\frac{\partial}{\partial t}\left(\frac{\partial D}{\partial \bar{u}_{j,t}}\right)=0, ~~ x=0,~~j=1,2,
 \label{NLSda}
 \\
\tilde{u}_{j,x}-\frac{\partial D}{\partial \bar{\tilde{u}}_j}+\frac{\partial}{\partial t}\left(\frac{\partial D}{\partial \bar{\tilde{u}}_{j,t}}\right)=0, ~~ x=0,~~j=1,2.
 \label{NLSdb}
\end{eqnarray}
\label{NLSd}
\end{subequations}
Equations (\ref{NLSa}) and (\ref{NLSb}) give nothing but the coupled NLS equation in the bulk regions $x>0$ and
$x<0$. We need to find a defect term $D$ that fits the defect conditions (\ref{NLSda}) and (\ref{NLSdb}).
Based on the BT (\ref{BTCNLS}) found for the coupled NLS equation,
we find that
\begin{eqnarray}
\begin{split}
D=&\frac{i}{2}\left[-3b\partial_t\ln\left(\frac{(u_1-\tilde{u}_1)(u_2-\tilde{u}_2)}{(\bar{u}_1-\bar{\tilde{u}}_1)(\bar{u}_2-\bar{\tilde{u}}_2)}\right)
+\frac{1}{\Omega+3b}\sum_{j=1}^2\left(|u_j-\tilde{u}_j|^2\partial_t\ln\left(\frac{u_j-\tilde{u}_j}{\bar{u}_j-\bar{\tilde{u}}_j}\right)\right)
\right]
\\&-\left(\frac{1}{3}\Omega^3+\Omega\left(|\tilde{u}_1|^2+|\tilde{u}_2|^2+|u_1|^2+|u_2|^2\right)
+\frac{|u_1\tilde{u}_2-\tilde{u}_1u_2|^2}{\Omega+3b}\right)
\\&-\left(3ia\left(u_1\bar{\tilde{u}}_1-\bar{u}_1\tilde{u}_1+u_2\bar{\tilde{u}}_2-\bar{u}_2\tilde{u}_2\right)-9a^2\Omega\right).
\end{split}
\label{DLD}
\end{eqnarray}
meets the requirement perfectly. With this choice,
the defect conditions (\ref{NLSda}) and (\ref{NLSdb}), after some algebra, give exactly the BT (\ref{BTCNLS}) frozen at $x=0$.

\subsection{Integrability of the defect system: Conserved quantities}
\begin{proposition}\label{defconservation}
For the defect coupled NLS equation defined above, a generating function of the conserved quantities is given by
\begin{eqnarray}
I(\lambda)=I_{bulk}^{left}(\lambda)+I_{bulk}^{right}(\lambda)+I_{defect}(\lambda),
\label{IM}
\end{eqnarray}
where
\begin{subequations}
\begin{eqnarray}
I_{bulk}^{left}(\lambda)=\int_{-\infty}^{0}\left(\tilde{u}_1\tilde{\Gamma}_1+\tilde{u}_2\tilde{\Gamma}_2 \right)dx,
\label{IMdefecta}
\\
I_{bulk}^{right}(\lambda)=\int_{0}^{\infty}\left(u_1\Gamma_1+u_2\Gamma_2 \right)dx,
\label{IMdefectb}
\\
I_{defect}(\lambda)=\left.\ln\left(B_{11}+B_{12}\tilde{\Gamma}_1+B_{13}\tilde{\Gamma}_2\right)\right|_{x=0}.
\label{IMdefectc}
\end{eqnarray}
\label{IMdefect}
\end{subequations}
In (\ref{IMdefectc}), $B_{jk}$, $1\leq j,k\leq3$, denote the $jk$-entries of the matrix $B(x,t,\lambda)$
(see (\ref{cnlsdm}) and (\ref{BTM}) for their explicit expressions).
\end{proposition}
{\bf Proof}
From (\ref{CL}), we obtain
\begin{eqnarray}
\begin{split}
-\left(\int_{-\infty}^{0}\left(\tilde{u}_1\tilde{\Gamma}_1+\tilde{u}_2\tilde{\Gamma}_2 \right)dx+\int_{0}^{\infty}\left(u_1\Gamma_1+u_2\Gamma_2 \right)dx\right)_t
\\
=\left.\left(V_{11}+V_{12}\Gamma_1+V_{13}\Gamma_2
-\tilde{V}_{11}-\tilde{V}_{12}\tilde{\Gamma}_1-\tilde{V}_{13}\tilde{\Gamma}_2
\right)\right|_{x=0},
\end{split}
\label{der1}
\end{eqnarray}
where we have used the vanishing boundary conditions for the fields $u_1(x,t)$, $u_2(x,t)$ as $x\rightarrow \infty$
and for the fields $\tilde{u}_1(x,t)$, $\tilde{u}_2(x,t)$ as $x\rightarrow -\infty$.
Equation (\ref{DV}) implies that the right hand side of (\ref{der1}) is the total time derivative of the functional (\ref{IMdefectc}).
Thus we have shown
\begin{eqnarray}
\frac{d\left(I(\lambda)\right)}{dt}=0.
\end{eqnarray}
This completes the proof. \QEDB

By inserting the expansion (\ref{Gamexpansion}) with (\ref{wj}) into (\ref{IM}),
we immediately obtain explicit forms of the conserved quantities.
For example, the first three conserved quantities are given by
\begin{eqnarray}
\begin{split}
I_1=&\int_{-\infty}^{0}\left(|\tilde{u}_1|^2+|\tilde{u}_2|^2\right)dx+\int_0^{\infty}\left(|u_1|^2+|u_2|^2\right)dx
-\left.\Omega\right|_{x=0},
\\
I_2=&\int_{-\infty}^{0}i\left(\tilde{u}_1\bar{\tilde{u}}_{1,x}-\bar{\tilde{u}}_1\tilde{u}_{1,x}+
\tilde{u}_2\bar{\tilde{u}}_{2,x}-\bar{\tilde{u}}_2\tilde{u}_{2,x}\right)dx
+\int_0^{\infty}i\left(u_1\bar{u}_{1,x}-\bar{u}_1u_{1,x}+u_2\bar{u}_{2,x}-\bar{u}_2u_{2,x}\right)dx
\\&+\left.i\left(u_1\bar{\tilde{u}}_1-\bar{u}_1\tilde{u}_1+u_2\bar{\tilde{u}}_2-\bar{u}_2\tilde{u}_2+6ia\Omega\right)\right|_{x=0},
\\
I_3=&\int_{-\infty}^{0}\left(\left(|\tilde{u}_1|^2+|\tilde{u}_2|^2\right)^2-|\tilde{u}_{1,x}|^2-|\tilde{u}_{2,x}|^2\right)dx+
\int_0^{\infty}\left(\left(|u_1|^2+|u_2|^2\right)^2-|u_{1,x}|^2-|u_{2,x}|^2\right)dx
\\&-\left.\left(\frac{1}{3}\Omega^3+\Omega\left(|\tilde{u}_1|^2+|\tilde{u}_2|^2+|u_1|^2+|u_2|^2\right)
-\frac{(\Omega-3b)|u_1\tilde{u}_2-\tilde{u}_1u_2|^2}{|\tilde{u}_1-u_1|^2+|\tilde{u}_2-u_2|^2}\right)\right|_{x=0}
\\&-\left.\left(3ia\left(u_1\bar{\tilde{u}}_1-\bar{u}_1\tilde{u}_1+u_2\bar{\tilde{u}}_2-\bar{u}_2\tilde{u}_2\right)
-9a^2\Omega\right)\right|_{x=0}.
\end{split}
\label{ecd}
\end{eqnarray}
The above expressions, up to suitable scaling, correspond to the modified norm, momentum and energy for the defect system, respectively.

\subsection{Integrability of the defect system: classical $r$-matrix approach}

The argument in section 3 implies that our type I defect condition can be interpreted simply as a canonical transformation with respect to the equal-space Poisson bracket (\ref{PBT}).
Here we further point out that the defect Lagrangian density (\ref{DLD}) found above can be reinterpreted as the density for the generating functional of the canonical transformation.
This provides us an explicit check that our type I defect condition is indeed a canonical transformation with respect to  the new equal-space Poisson structure.
To show this, we first note that the coupled NLS equation in equal-space Hamiltonian formulation corresponds to the following Pfaffian form, that is relative integrable invariant,
\begin{eqnarray}
\theta=\int_{-\infty}^{\infty}dt\left(\bar{u}_{1,x}du_1+u_{1,x}d\bar{u}_1+\bar{u}_{2,x}du_2+u_{2,x}d\bar{u}_2\right)-H_Tdx,
\label{pffi1}
\end{eqnarray}
where $d$ denotes an exterior derivative defined on the functional space of states $(u_1,u_2,\bar{u}_1,\bar{u}_2,x)$.
Indeed, the exterior differential form $d\theta$ is given by
\begin{eqnarray}
\begin{split}
d\theta=\int_{-\infty}^{\infty}dt\{&d\bar{u}_{1,x}\wedge du_1+du_{1,x}\wedge d\bar{u}_1+d\bar{u}_{2,x}\wedge du_2+du_{2,x}\wedge d\bar{u}_2
\\
&-\sum_{j=0}(-1)^j\frac{\partial^j}{\partial t^j}\frac{\partial\mathcal{H}_T}{\partial u_{1,t_j}}du_1\wedge dx
-\sum_{j=0}(-1)^j\frac{\partial^j}{\partial t^j}\frac{\partial\mathcal{H}_T}{\partial u_{2,t_j}}du_2\wedge dx
\\
&-\sum_{j=0}(-1)^j\frac{\partial^j}{\partial t^j}\frac{\partial\mathcal{H}_T}{\partial \bar{u}_{1,t_j}}d\bar{u}_1\wedge dx
-\sum_{j=0}(-1)^j\frac{\partial^j}{\partial t^j}\frac{\partial\mathcal{H}_T}{\partial \bar{u}_{2,t_j}}d\bar{u}_2\wedge dx\},
\end{split}
\label{dpffi1}
\end{eqnarray}
where $u_{1,t_j}\equiv \frac{\partial^j}{\partial t^j}u_1$, $u_{2,t_j}\equiv \frac{\partial^j}{\partial t^j}u_2$, $j=0,1,2,\cdots$, and the symbol $\wedge$ denotes an exterior product.
If the one-form $\theta$ is a relative integrable invariant, then we obtain
\begin{eqnarray}
u_{1,xx}=-\sum_{j=0}(-1)^j\frac{\partial^j}{\partial t^j}\frac{\partial\mathcal{H}_T}{\partial \bar{u}_{1,t_j}},
~~u_{2,xx}=-\sum_{j=0}(-1)^j\frac{\partial^j}{\partial t^j}\frac{\partial\mathcal{H}_T}{\partial \bar{u}_{2,t_j}},
\end{eqnarray}
which is exactly the coupled NLS equation in the time Hamiltonian form.
Transformation, which maps $\left(u_1,u_2\right)$ to $\left(\tilde{u}_1,\tilde{u}_2\right)$, is canonical if the Pfaffian form (\ref{pffi1}) is relative integrable invariant under the transformation. That is
\begin{eqnarray}
\begin{split}
&\int_{-\infty}^{\infty}dt\left(\bar{\tilde{u}}_{1,x}d\tilde{u}_1+\tilde{u}_{1,x}d\bar{\tilde{u}}_1
+\bar{\tilde{u}}_{2,x}d\tilde{u}_2+\tilde{u}_{2,x}d\bar{\tilde{u}}_2\right)-\tilde{H}_Tdx
\\=&\int_{-\infty}^{\infty}dt\left(\bar{u}_{1,x}du_1+u_{1,x}d\bar{u}_1+\bar{u}_{2,x}du_2+u_{2,x}d\bar{u}_2\right)-H_Tdx+dW,
\end{split}
\label{pffi2}
\end{eqnarray}
where
\begin{eqnarray}
W=\int_{-\infty}^{\infty}dt\mathcal{F}(u_1,u_2,\bar{u}_1,\bar{u}_2,\tilde{u}_1,\tilde{u}_2,\bar{\tilde{u}}_1,\bar{\tilde{u}}_2)+Ex
\end{eqnarray}
(with $E$ being a real constant) is the so-called generator of the transformation.
Equation (\ref{pffi2}) implies the following transformation formulae:
\begin{eqnarray}
\begin{split}
u_{1,x}=-\sum_{j=0}(-1)^j\frac{\partial^j}{\partial t^j}\frac{\partial\mathcal{F}}{\partial \bar{u}_{1,t_j}},
~~u_{2,x}=-\sum_{j=0}(-1)^j\frac{\partial^j}{\partial t^j}\frac{\partial\mathcal{F}}{\partial \bar{u}_{2,t_j}},
\\
\tilde{u}_{1,x}=\sum_{j=0}(-1)^j\frac{\partial^j}{\partial t^j}\frac{\partial\mathcal{F}}{\partial \bar{\tilde{u}}_{1,t_j}},
~~\tilde{u}_{2,x}=\sum_{j=0}(-1)^j\frac{\partial^j}{\partial t^j}\frac{\partial\mathcal{F}}{\partial \bar{\tilde{u}}_{2,t_j}}.
\end{split}
\label{Transfe}
\end{eqnarray}
By choosing $\mathcal{F}=D$, where $D$ is given by (\ref{DLD}), we find that (\ref{Transfe}), after some algebra, are exactly the BT (\ref{BTCNLS}).

The canonical property of the defect condition enables us to implement the classical $r$-matrix approach to establish the Liouville integrability of the defect coupled NLS system from the point of view of the equal-space Poisson structure.
The argument follows the analogous argument in the context of the NLS equation \cite{CK2015}. It goes as follows.
For the system in the bulk region $x>0$, we define the transition matrix by $M_T(t,\tau,\lambda)$, where $M_T(t,\tau,\lambda)$ is the matrix introduced in section 2.2.
For the system in the bulk region $x<0$, we define the transition matrix by $\widetilde{M}_T(t,\tau,\lambda)$,
where $\widetilde{M}_T(t,\tau,\lambda)$ is analogous to $M_T(t,\tau,\lambda)$ but constructed from the new canonical variables.
At the defect location $x=0$, we change the variables describing the system from the old canonical ones to the new canonical ones, thus we have two equivalent options to represent the transition matrix, that is $\widetilde{M}_T(t,\tau,\lambda)$ or $M_T(t,\tau,\lambda)$
(they are connected by $\widetilde{M}_T(t,\tau,\lambda)=B(x,t,\lambda)M_T(t,\tau,\lambda)$ at $x=0$).
In summary, we construct the transition matrix for the defect system as follows
\begin{eqnarray}
\begin{split}
\mathcal{M}(x,t,\tau,\lambda)=\left\{\begin{array}{l}
\widetilde{M}_T(t,\tau,\lambda), \quad -\infty<x<0,
\\
M_T(t,\tau,\lambda), \quad 0\leq x<\infty.
\end{array}\right.
\end{split}
\label{M1}
\end{eqnarray}
We immediately conclude that $\mathcal{M}(x,t,\tau,\lambda)$ satisfies the same $r$-matrix relation as that of $M_T(t,\tau,\lambda)$, that is
\begin{eqnarray}
\left\{\mathcal{M}_{1}(x,t,\tau,\lambda),\mathcal{M}_{2}(x,t,\tau,\mu)\right\}_T=\left[r(\lambda-\mu),\mathcal{M}(x,t, \tau,\lambda)\otimes \mathcal{M}(x,t,\tau,\mu)\right].
\label{rmatreM}
\end{eqnarray}
As a result, the trace of the monodromy matrix $\mathcal{M}(x,\infty,-\infty,\lambda)$ generates the conserved quantities that are in involution with respect to the Poisson bracket (\ref{PBT}).


\section{New integrable boundary conditions for the coupled NLS equation}

In this section, we will study new integrable boundary conditions associated with the coupled NLS equation.
Our new boundary conditions are derived by imposing suitable reductions on the defect condition, that is the BT  (\ref{BTCNLS}) frozen at $x=0$. It is worth reminding that the idea to construct integrable boundary conditions via a BT was initiated by Habibullin \cite{Habibullin1991}.
The main difference with respect to the method of Habibullin is that in our method we exploit simultaneously the space-part and the time-part of the BT, whereas, in \cite{Habibullin1991}, only the space-part of the BT is considered. As we will show below the time-part of the BT, indeed, yields new integrable boundary conditions and it is very instructive for us to prove the integrability of the corresponding boundary conditions.

\subsection{New boundary conditions arising from $U(2)$ symmetry reductions of the defect condition}

One can check directly that if $\left(u_1(x,t),u_2(x,t)\right)^T$ solves the coupled NLS equation (\ref{CNLS}), then so does $C \left(u_1(-x,t),u_2(-x,t)\right)^T$, where $C$ is a $2\times 2$ unitary matrix, that is $C$ satisfies $C^{\dagger}C=CC^{\dagger}=I_2$.
This fact implies that we may consider the following reductions between the fields $\left(u_1,u_2\right)$ and $\left(\tilde{u}_1,\tilde{u}_2\right)$,
\begin{eqnarray}
\left(\tilde{u}_1(x,t),\tilde{u}_2(x,t)\right)^T=C \left(u_1(-x,t),u_2(-x,t)\right)^T,
\label{reduction}
\end{eqnarray}
where $C\equiv (c_{jk})_{2\times 2}$ is a $2\times 2$ unitary matrix as before.
By imposing this reduction on (\ref{BTCNLS}) and evaluating the resulting equations at $x=0$,
we obtain the following boundary condition for the coupled NLS equation
\begin{eqnarray}
\begin{split}
(1+c_{11})u_{1,x}(0)+c_{12}u_{2,x}(0)&=-3iB^{(0)}_{11}(0)\sum_{j=1}^2c_{1j}u_j(0)
+3iu_1(0)B^{(0)}_{22}(0)+3iu_2(0)B^{(0)}_{32}(0),
\\
c_{21}u_{1,x}(0)+(1+c_{22})u_{2,x}(0)&=-3iB^{(0)}_{11}(0)\sum_{j=1}^2c_{2j}u_j(0)
+3iu_1(0)B^{(0)}_{23}(0)+3iu_2(0)B^{(0)}_{33}(0),
\\
(1-c_{11})u_{1,t}(0)-c_{12}u_{2,t}(0)&=2i\left(\left(1-c_{11}\right)u_1(0)-c_{12}u_2(0)\right)
\left(|u_1(0)|^2+|u_2(0)|^2\right)
\\&+i\left(\left(c_{11}-c_{22}\right)u_1(0)u_2(0)+c_{12}u^2_2(0)-c_{21}u^2_1(0)\right)
\sum_{j=1}^2\bar{c}_{2j}\bar{u}_j(0)
\\&-3B^{(0)}_{11}(0)\sum_{j=1}^2c_{1j}u_{j,x}(0)
-3u_{1,x}(0)B^{(0)}_{22}(0)-3u_{2,x}(0)B^{(0)}_{32}(0),
\\
(1-c_{22})u_{2,t}(0)-c_{21}u_{1,t}(0)&=2i\left(\left(1-c_{22}\right)u_2(0)-c_{21}u_1(0)\right)
\left(|u_1(0)|^2+|u_2(0)|^2\right)
\\&+i\left(\left(c_{22}-c_{11}\right)u_1(0)u_2(0)+c_{21}u^2_1(0)-c_{12}u^2_2(0)\right)
\sum_{j=1}^2\bar{c}_{1j}\bar{u}_j(0)
\\&-3B^{(0)}_{11}(0)\sum_{j=1}^2c_{2j}u_{j,x}(0)
-3u_{1,x}(0)B^{(0)}_{23}(0)-3u_{2,x}(0)B^{(0)}_{33}(0),
\end{split}
\label{gubc}
\end{eqnarray}
where $B^{(0)}_{jk}(0)$, $1\leq j,k\leq 3$, are the quantities that obtained by inserting (\ref{reduction}) into the  quantities (\ref{BTM}) evaluated at $x=0$.
We will show in section 5.3 that the boundary condition (\ref{gubc}) in general is not integrable,
except for the following three cases.

{\bf Case 1: Robin boundary condition corresponding to (\ref{gubc}) with $C=\diag(1,1)$ and $a=0$.}
In this case, the reduction (\ref{reduction}) reads
\begin{eqnarray}
\tilde{u}_1(x,t)=u_1(-x,t),~~\tilde{u}_2(x,t)=u_2(-x,t).
\label{Red1}
\end{eqnarray}
and the boundary condition (\ref{gubc}) becomes
\begin{eqnarray}
\left(u_{1,x}-\alpha u_1\right)\mid_{x=0}=0, ~~\left(u_{2,x}-\alpha u_2\right)\mid_{x=0}=0,
\label{RBC}
\end{eqnarray}
where $\alpha=3b$ is a real constant.
This boundary condition is the two-component generalisation of the usual Robin boundary condition in the scalar case.
As limiting cases, we recover the Dirichlet boundary condition
$u_1\mid_{x=0}= u_2\mid_{x=0}=0$ as $\alpha\rightarrow \infty$, and recover the Neumann boundary condition
$u_{1,x}\mid_{x=0}= u_{2,x}\mid_{x=0}=0$ as $\alpha\rightarrow 0$.
We note that the vector generalisation of the usual Robin boundary condition for the half-line problem has been studied in \cite{CZ2012,ZZ2021}.

{\bf Case 2: new boundary condition corresponding to (\ref{gubc}) with $C=\diag(-1,-1)$ and $a=0$.}
In this case, the reduction (\ref{reduction}) reads
\begin{eqnarray}
\tilde{u}_1(x,t)=-u_1(-x,t),~~\tilde{u}_2(x,t)=-u_2(-x,t),
\label{Red2}
\end{eqnarray}
and the boundary condition (\ref{gubc}) becomes
\begin{subequations}
\begin{eqnarray}
\left.\left(iu_{1,t}+2 u_1\left(|u_1|^2+|u_2|^2\right)+u_{1,x}\hat{\Omega}+
2\frac{u_{1,x}|u_2|^2-u_{2,x}u_1\bar{u}_2}{\hat{\Omega}+3b}\right)\right|_{x=0}=0,
\label{tbca}
\\
\left.\left(iu_{2,t}+2 u_2\left(|u_1|^2+|u_2|^2\right)+u_{2,x}\hat{\Omega}+
2\frac{u_{2,x}|u_1|^2-u_{1,x}\bar{u}_1u_2}{\hat{\Omega}+3b}\right)\right|_{x=0}=0,
\label{tbcb}
\end{eqnarray}
\label{tbc}
\end{subequations}
where
\begin{eqnarray}
\hat{\Omega}\equiv\hat{\Omega}(x,t)=\sqrt{9b^2-4\left(|u_1|^2+|u_2|^2\right)}.
\label{hatomega}
\end{eqnarray}
Unlike the Robin boundary condition, this new boundary condition is non-linear, and it involves time derivatives of the coupled NLS fields.

{\bf Case 3: new boundary conditions corresponding to (\ref{gubc}) with $a=0$ and with $C$ satisfying
$\tr C=0$, $C=C^{\dagger }$ and $\det C=-1$.}
In this case, the reduction becomes
\begin{eqnarray}
\left( \begin{array}{cc} \tilde{u}_1(x,t) \\
 \tilde{u}_2(x,t) \\ \end{array} \right)=\left( \begin{array}{cc} c_{11} & c_{12} \\
 \bar{c}_{12} &  -c_{11} \\ \end{array} \right)
 \left( \begin{array}{cc} u_1(-x,t) \\
 u_2(-x,t) \\ \end{array} \right),~~c_{11}\in \mathbb{R},~~c^2_{11}+|c_{12}|^2=1.
\label{Red3g}
\end{eqnarray}
In particular, for $c_{11}=1$ and $c_{12}=0$,
the reduction becomes
\begin{eqnarray}
\tilde{u}_1(x,t)=u_1(-x,t),~~\tilde{u}_2(x,t)=-u_2(-x,t),
\label{Red3}
\end{eqnarray}
and the associated boundary condition reads
\begin{subequations}
\begin{eqnarray}
\left.\left(2u_{1,x}- u_1(\Omega_2+3b)\right)\right|_{x=0}=0,
\label{bcase3a}
 \\
\left.\left(iu_{2,t}+ u_2\left(|u_1|^2+2|u_2|^2\right)+u_{2,x}\Omega_2\right)\right|_{x=0}=0,
\label{bcase3b}
\end{eqnarray}
\label{bcase3}
\end{subequations}
where
\begin{eqnarray}
\Omega_2\equiv \Omega_2(x,t)=\sqrt{9b^2-4|u_2|^2}.\label{Omg2}
\end{eqnarray}
We note that the boundary condition corresponding to the general reduction (\ref{Red3g}),
after being rewritten in terms of the variables
$$\hat{u}_1=\frac{1}{\sqrt{2(1+c_{11})}}\left((1+c_{11})u_1+c_{12} u_2\right),~~
\hat{u}_2=\frac{1}{\sqrt{2(1+c_{11})}}\left(-\bar{c}_{12} u_1+(1+c_{11})u_2\right),$$
is equal to the boundary condition (\ref{bcase3}).
Thus, it is enough for us to consider the boundary condition (\ref{bcase3}) in this case.

\subsection{A new boundary condition derived via dressing the Dirichlet boundary with the integrable defect condition}

For the scalar NLS equation, it was found in \cite{Zambon2014} that dressing the Dirichlet boundary with a type I defect generates a new boundary, which is
\begin{eqnarray}
\left.\left(iu_t+2u_{x}\sqrt{4b^2-|u|^2}-4(a^2+b^2)u+2 u|u|^2\right)\right|_{x=0}=0.
\label{tbcNLS}
\end{eqnarray}
Here we generalise this result to the coupled case.

We assume that the fields $\tilde{u}_1$ and $\tilde{u}_2$ at the boundary satisfies the Dirichlet boundary condition, that is $$\left.\tilde{u}_{1}\right|_{x=0}=\left.\tilde{u}_{2}\right|_{x=0}.$$
Inserting this condition to the defect condition, we find, at $x=0$,
\begin{subequations}
\begin{eqnarray}
u_{1,x}-\tilde{u}_{1,x}=u_1\left(\Omega_{12}+3ia\right),
\label{BTNLSr1}
\\
u_{2,x}-\tilde{u}_{2,x}=u_2\left(\Omega_{12}+3ia\right),
\label{BTNLSr2}
\\
u_{1,t}=iu_1(|u_1|^2+|u_2|^2)+i\tilde{u}_{1,x}(\Omega_{12}-3ia)-3u_{1,x}(a-ib)
+\frac{u_{1,x}|u_1|^2+u_{2,x}u_1\bar{u}_2}{i(\Omega_{12}+3b)},
\label{BTNLSr3}
\\
u_{2,t}=iu_2(|u_1|^2+|u_2|^2)+i\tilde{u}_{2,x}(\Omega_{12}-3ia)-3u_{2,x}(a-ib)
+\frac{u_{1,x}\bar{u}_1u_2+u_{2,x}|u_2|^2}{i(\Omega_{12}+3b)}.
\label{BTNLSr4}
\end{eqnarray}
\label{BTNLSr}
\end{subequations}
where
\begin{eqnarray}
\Omega_{12}\equiv \Omega_{12}(x,t)=\sqrt{9b^2-|u_1|^2-|u_2|^2}.
\label{omg12}
\end{eqnarray}
By using (\ref{BTNLSr1}) and (\ref{BTNLSr2}) to eliminate $\tilde{u}_{1,x}$ and $\tilde{u}_{2,x}$ in (\ref{BTNLSr3}) and (\ref{BTNLSr4}), we find the following new boundary condition
\begin{eqnarray}
\begin{split}
\left.\left(iu_{1,t}+ u_1\left(2(|u_1|^2+|u_2|^2)-9(a^2+b^2)\right)+2u_{1,x}\Omega_{12}+
\frac{u_{1,x}|u_2|^2-u_{2,x}u_1\bar{u}_2}{\Omega_{12}+3b}\right)\right|_{x=0}=0,
\\
\left.\left(iu_{2,t}+ u_2\left(2(|u_1|^2+|u_2|^2)-9(a^2+b^2)\right)+2u_{2,x}\Omega_{12}+
\frac{u_{2,x}|u_1|^2-u_{1,x}\bar{u}_1u_2}{\Omega_{12}+3b}\right)\right|_{x=0}=0.
\end{split}
\label{bcase5}
\end{eqnarray}
This new boundary condition is a two-component generalization of the Zambon's new boundary condition (\ref{tbcNLS}).

\subsection{Integrability of the boundary conditions: $r$-matrix approach}\label{sec5.2}

We now study the integrability of our new boundary conditions by using the $r$-matrix approach.
Following Sklyanin's formalism \cite{Sklyanin1987}, in order to study the integrability of a boundary problem on the half-line,
it is important to consider the following generalization of the monodromy matrix
\begin{eqnarray}
\mathcal{M}(\lambda)=M(\lambda)K(\lambda)M^{-1}(-\lambda),
\label{gm}
\end{eqnarray}
where $M(\lambda)=M_S(\infty,0,\lambda)$.
We assume, in general, the $K(\lambda)$ matrix can depend on time \cite{ACC2018}.
One has the following results, which can be proved by direct computations.
\begin{lemma}
If $K(\lambda)$ satisfies relations
\begin{eqnarray}
\begin{split}
\left\{K_1(\lambda),K_2(\mu)\right\}=&\left[r(\lambda-\mu),K_1(\lambda)K_2(\mu)\right]
+K_1(\lambda)r(\lambda+\mu)K_2(\mu)
\\&-K_2(\mu)r(\lambda+\mu)K_1(\lambda)
\end{split}
 \label{alga}
\end{eqnarray}
and
\begin{eqnarray}
\left\{K_1(\lambda),U_2(x,t,\mu)\right\} =0,
 \label{algb}
\end{eqnarray}
 then
\begin{eqnarray}
\begin{split}
\left\{\mathcal{M}_1(\lambda),\mathcal{M}_2(\mu)\right\}=
&\left[r(\lambda-\mu),\mathcal{M}_1(\lambda)\mathcal{M}_2(\mu)\right]
\\&+\mathcal{M}_1(\lambda)r(\lambda+\mu)\mathcal{M}_2(\mu)-\mathcal{M}_2(\mu)r(\lambda+\mu)\mathcal{M}_1(\lambda).
\end{split}
\label{Mr}
\end{eqnarray}
\end{lemma}
\begin{lemma}
If the $K(\lambda)$ matrix satisfies the following equation at the boundary
\begin{eqnarray}
 \frac{dK(\lambda)}{dt}=V(0,t,\lambda)K(\lambda)-K(\lambda)V(0,t,-\lambda),
 \label{algc}
\end{eqnarray}
then
 \begin{eqnarray}
 \frac{d\tr(\mathcal{M}(\lambda))}{dt}=0.
 \end{eqnarray}
\end{lemma}
Lemma 2 implies that the quantity $\tr(\mathcal{M}(\lambda))$ Poisson
commutes with itself for different values of the spectral parameter, if the $K(\lambda)$ matrix subjects to the  boundary Poisson algebra (\ref{alga}) and (\ref{algb}).
Lemma 3 enables us to interpret the quantity $\tr(\mathcal{M}(\lambda))$ as the generating function of the conserved quantities.
Thus, for the coupled NLS equation with a boundary resulting from (\ref{algc}), the integrability is obtained.

The above argument implies that the integrable boundary conditions are encoded into the boundary $K(\lambda)$ matrices that satisfy equations (\ref{alga}), (\ref{algb}) and (\ref{algc}).
In order to establish the integrability of our new boundary conditions, we need to find the corresponding $K(\lambda)$ matrices.
In general, it is not easy to find a $K(\lambda)$ matrix that matches the boundary equation (\ref{algc}).
Here we present a connection between the BT matrices and the solutions of the boundary equation (\ref{algc}).
Based on this, we are able to derive the $K(\lambda)$ matrices for all the boundary conditions presented in sections 5.1 and 5.2.
\begin{proposition}\label{pro2}
Let the fields $(\tilde{u}_1, \tilde{u}_2)$ and $(u_1, u_2)$ subject to a suitable reduction
(say (\ref{reduction}) for example).
If, under this reduction, there exists a non-degenerate and time-independent matrix $P(\lambda)$ such that
\begin{eqnarray}
\tilde{V}(0,t,\lambda)=P(\lambda)V(0,t,-\lambda)P^{-1}(\lambda),
\label{VP}
\end{eqnarray}
then
\begin{eqnarray}
K(\lambda)=B(0,t,\lambda)P(\lambda),
\label{KB}
\end{eqnarray}
where $B(0,t,\lambda)$ is the BT matrix evaluated at $x=0$,
satisfies the boundary equation (\ref{algc}).
\end{proposition}
{\bf Proof}
Evaluating the time-part of the BT (\ref{BT1b}) at $x=0$ and using (\ref{VP}), we obtain
\begin{eqnarray}
B_t(0,t,\lambda)=V(0,t,\lambda)B(0,t,\lambda)-B(0,t,\lambda)P(\lambda)V(0,t,-\lambda)P^{-1}(\lambda),
\label{BT1bVP}
\end{eqnarray}
which can be written as
\begin{eqnarray}
\left(B(0,t,\lambda)P(\lambda)\right)_t=V(0,t,\lambda)B(0,t,\lambda)P(\lambda)-B(0,t,\lambda)P(\lambda)V(0,t,-\lambda),
\label{BT1bVP1}
\end{eqnarray}
where the time-independent property of $P(\lambda)$ matrix is used.
By comparing (\ref{BT1bVP1}) with (\ref{algc}), we complete the proof.
\QEDB

For the $U(2)$ symmetry reduction (\ref{reduction}), we find that the $P(\lambda)$ matrix in (\ref{VP}) is given by
\begin{eqnarray}
P=\left( \begin{array}{ccc}
1 & 0 & 0 \\
 0 &  -\bar{c}_{11} & -\bar{c}_{12} \\
 0 &  -\bar{c}_{21} & -\bar{c}_{22} \\
 \end{array} \right),
\label{P}
\end{eqnarray}
where $c_{jk}$, $1\leq j,k\leq 2$, are $jk$-entries of the unitary matrix $C$.
Thus the $K(\lambda)$ matrix corresponding to the general boundary condition (\ref{gubc}) is given by
\begin{eqnarray}
\begin{split}
K(\lambda)=\left.\mathcal{B}(x,t,\lambda)\right|_{x=0}P\equiv P+\lambda^{-1}K^{(0)},
\end{split}
\label{K}
\end{eqnarray}
where $K^{(0)}=\left.\mathcal{B}^{(0)}(x,t)\right|_{x=0}P$, and $\mathcal{B}$ and $\mathcal{B}^{(0)}$ stand for the matrices that obtained by inserting the reduction (\ref{reduction}) into the matrices $B$ and $B^{(0)}$ (see (\ref{cnlsdm})), respectively.
For example, for the Robin boundary condition (\ref{RBC}), the corresponding $K(\lambda)$ matrix is given by
\begin{eqnarray}
K(\lambda)=\diag\left(1+ib\lambda^{-1},~-1+ib\lambda^{-1},~-1+ib\lambda^{-1}\right).
\label{K1}
\end{eqnarray}
For the new boundary condition (\ref{tbc}), the corresponding $K(\lambda)$ matrix is given by
\begin{eqnarray}
\begin{split}
K(\lambda)
= I_3+\lambda^{-1}\left( \begin{array}{ccc}
\frac{i}{3}\hat{\Omega}(0) & -\frac{2i}{3}u_1(0) & -\frac{2i}{3}u_2(0) \\
 -\frac{2i}{3}\bar{u}_1(0) &  -ib+\frac{4i|u_1(0)|^2}{3(\hat{\Omega}(0)+3b)} & \frac{4i\bar{u}_1(0)u_2(0)}{3(\hat{\Omega}(0)+3b)} \\
 -\frac{2i}{3}\bar{u}_2(0) & \frac{4iu_1(0)\bar{u}_2(0)}{3(\hat{\Omega}(0)+3b)} & -ib+\frac{4i|u_2(0)|^2}{3(\hat{\Omega}(0)+3b)} \\
 \end{array} \right)
 \equiv I_3+\lambda^{-1} K^{(0)},
 \label{K2}
 \end{split}
\end{eqnarray}
where we have used the notations $u_1(0)=u_1(0,t)$, $u_2(0)=u_2(0,t)$, $\hat{\Omega}(0)=\hat{\Omega}(0,t)$.
For the new boundary condition (\ref{bcase3}), the corresponding $K(\lambda)$ matrix is given by
\begin{eqnarray}
K(\lambda)=P+\lambda^{-1}\left( \begin{array}{ccc}
\frac{i}{3}\Omega_2(0) & 0 & -\frac{2i}{3}u_2(0) \\
 0 &  ib & 0 \\
 -\frac{2i}{3}\bar{u}_2(0) & 0 & -\frac{i}{3}\Omega_2(0)  \\
 \end{array} \right)
 \equiv P+\lambda^{-1} K^{(0)},
 \label{K3}
\end{eqnarray}
where
\begin{eqnarray}
P=\diag\left(1,~-1,~1\right),~~\Omega_2(0)=\Omega_2(0,t).
\label{P3}
\end{eqnarray}
In comparison with the $K(\lambda)$ matrix (\ref{K1}) for the Robin boundary condition, the $K(\lambda)$ matrices (\ref{K2}) and (\ref{K3}) are no more constant matrices, they depend on the fields of the coupled NLS equation.

The above analysis shows that the general boundary conditions (\ref{gubc}) can be written in the form of (\ref{algc})
with the $K(\lambda)$ matrix given by (\ref{K}). Since (\ref{K}) holds at the boundary location, the Poisson bracket (\ref{algb}) is automatically zero. However, we should emphasize that the $K(\lambda)$ matrix (\ref{K}) in general does not match (\ref{alga}). Thus the boundary condition (\ref{gubc}) is not integrable in the general case.
We need to select suitable $C$ matrices such that the associated $K(\lambda)$ matrices match (\ref{alga}), and thus the resulting boundary conditions are integrable.
Using (\ref{K}), the left hand side of (\ref{alga}) becomes
\begin{eqnarray}
\frac{1}{\lambda\mu}\left\{K^{(0)}_1,K^{(0)}_2\right\},
\label{LHSalga}
\end{eqnarray}
where $K^{(0)}_1=K^{(0)}\otimes I_3$ and $K^{(0)}_2=I_3 \otimes K^{(0)}$.
After straightforward calculations using the formula (\ref{Prel}), the right hand side of (\ref{alga}) becomes
\begin{eqnarray}
\begin{split}
&\frac{1}{3\lambda\mu}\left[\mathcal{P},P\otimes K^{(0)}\right]
\\&+\frac{1}{3\left(\lambda+\mu\right)}\left\{\frac{1}{\lambda}\left[(K^{(0)}P)\otimes I_3-I_3\otimes(PK^{(0)}) \right]+\frac{1}{\mu}\left[(PK^{(0)})\otimes I_3-I_3\otimes(K^{(0)}P) \right]\right\}\mathcal{P}
\\
&+\frac{1}{3\left(\lambda+\mu\right)}\left\{P^2\otimes I_3-I_3\otimes P^2+\frac{1}{\lambda\mu}\left[(K^{(0)})^2\otimes I_3-I_3\otimes (K^{(0)})^2\right]\right\}\mathcal{P}.
\end{split}
\label{RHSalga}
\end{eqnarray}
In order to match (\ref{RHSalga}) with (\ref{LHSalga}),
the two terms $P^2\otimes I_3-I_3\otimes P^2$ and $(K^{(0)})^2\otimes I_3-I_3\otimes (K^{(0)})^2$
in the last line of the expression (\ref{RHSalga}) should vanish.
By requiring $P^2\otimes I_3-I_3\otimes P^2=0$, we find the entries of the matrix $C$ should satisfy either of the following two constraints
\begin{subequations}
\begin{eqnarray}
&&c_{11}=c_{22}=\pm 1,~~c_{12}=c_{21}=0,
\label{C1}
\\
&&c_{11}=-c_{22},~~\det C=-1.
\label{C2}
\end{eqnarray}
\label{C}
\end{subequations}
By requiring $(K^{(0)})^2\otimes I_3-I_3\otimes (K^{(0)})^2=0$, we find that, in addition to (\ref{C}), the entries of the unitary matrix $C$ should also obey $c_{11}=\bar{c}_{11}$ and $c_{12}=\bar{c}_{21}$, and we find that the parameter $a=0$. Under these constraints, (\ref{RHSalga}) is reduced to
\begin{eqnarray}
\begin{split}
&\frac{1}{3\lambda\mu}\left[\mathcal{P},P\otimes K^{(0)}+I_3\otimes(PK^{(0)})\right],
\end{split}
\label{RHSalga2}
\end{eqnarray}
which is possible to match (\ref{LHSalga}).
Combining the above analysis, we obtain the following possible solutions for the unitary matrix $C$,
\begin{subequations}
\begin{eqnarray}
c_{11}=c_{22}=1,~~c_{12}=c_{21}=0,
\\
c_{11}=c_{22}=-1,~~c_{12}=c_{21}=0,
\\
c_{11}=-c_{22},~~c_{11}=\bar{c}_{11},~~c_{12}=\bar{c}_{21},~~\det C=-1.
\end{eqnarray}
\end{subequations}
The boundary conditions corresponding to the above $C$ matrices are just the three ones that presented in section 5.1. We next show these boundary conditions are indeed integrable.

For the Robin boundary condition, the associated $K(\lambda)$ matrix (\ref{K1}) is non-dynamic, and thus the Poisson bracket (\ref{alga}) automatically holds. This confirms the integrability of the coupled NLS equation in the presence of the Robin boundary condition (\ref{RBC}).
However, for the new boundary conditions (\ref{tbc}) and (\ref{bcase3}), the situation is quite different:  the associated $K(\lambda)$ matrices (\ref{K2}) and (\ref{K3}) are dynamic. We need to introduce suitable Poisson brackets at the boundary in order to calculate the Poisson brackets of these dynamic $K(\lambda)$ matrices (the left hand side of (\ref{alga})).
For the boundary condition (\ref{tbc}), we introduce the following Poisson brackets at the boundary
\begin{eqnarray}
\begin{split}
\left\{u_1(0),\bar{u}_1(0)\right\}&=-\frac{3}{2}\left(K^{(0)}_{11}-K^{(0)}_{22}\right),
~\left\{u_1(0),\bar{u}_2(0)\right\}=\frac{3}{2}K^{(0)}_{32},
\\
\left\{u_2(0),\bar{u}_2(0)\right\}&=-\frac{3}{2}\left(K^{(0)}_{11}-K^{(0)}_{33}\right),
~\left\{u_2(0),\bar{u}_1(0)\right\}=\frac{3}{2}K^{(0)}_{23},
\\
\left\{u_1(0),u_2(0)\right\}&=\left\{\bar{u}_1(0),\bar{u}_2(0)\right\}=0,
\end{split}
\label{PBBK2}
\end{eqnarray}
where the quantities $K^{(0)}_{jk}$, $1 \leq j,k \leq 3$, are the $jk$-entries of the matrix $K^{(0)}$ that is defined in the expression (\ref{K2}).
After straightforward calculation using (\ref{K2}) and (\ref{PBBK2}), we find that (\ref{LHSalga}) is equal to (\ref{RHSalga2}), as expected. As a consequence, for the $K(\lambda)$ matrix (\ref{K2}), expression (\ref{alga}) becomes an identity. Thus integrability of the new boundary condition (\ref{tbc}) is obtained.
For the new boundary condition (\ref{bcase3}), we introduce the following boundary Poisson bracket
\begin{eqnarray}
\begin{split}
\left\{u_2(0),\bar{u}_2(0)\right\}=-i\Omega_2(0).
\end{split}
\label{PBBK3}
\end{eqnarray}
By using this Poisson bracket and (\ref{K3}), we can check directly that (\ref{LHSalga}) is equal to (\ref{RHSalga2}). Thus, for the $K(\lambda)$ matrix (\ref{K3}), the expression (\ref{alga})  becomes an identity.
This implies the integrability of the new boundary condition (\ref{bcase3}).

We now turn to the integrability of the new boundary condition (\ref{bcase5}).
We first need to derive a $K(\lambda)$ matrix for the boundary condition (\ref{bcase5}).
As mentioned above, the new boundary condition (\ref{bcase5}) is a two component generalization of the boundary condition (\ref{tbcNLS}).
For the latter boundary condition (boundary (\ref{tbcNLS}) in the scalar case), it has been found that the associated $K(\lambda)$ matrix is of order $2$ in $\lambda^{-1}$ \cite{Zambon2014}, and it can be constructed as a composition of several BT matrices of order $1$ in $\lambda^{-1}$ \cite{CCD2021}.
This result implies that the structure of the $K(\lambda)$ matrix for the new boundary condition (\ref{bcase5}) should be different from the ones for the boundary conditions presented in section 5.1; it should also be of order $2$ in $\lambda^{-1}$ rather than order $1$.
We will show in the following it is indeed the case, and such a boundary condition can be constructed by applying proposition 2 to a BT matrix produced as a product of two BT matrices of the form (\ref{cnlsdm}).

We consider two BTs as follows.
The first one induced by a matrix $B^{(1)}$ of the form (\ref{cnlsdm}) with the parametres $a_1$ and $b_1$ produces $u^{(1)}_1$ and $u^{(1)}_2$ from $u^{(0)}_1$ and $u^{(0)}_2$. The second one induced by a matrix $B^{(2)}$ of the form (\ref{cnlsdm}) with the parametres $a_2$ and $b_2$ produces $u^{(2)}_1$ and $u^{(2)}_2$ from $u^{(1)}_1$ and $u^{(1)}_2$. More precisely, the two BT matrices $B^{(j)}$, $j=1,2$, are given by
\begin{eqnarray}
\begin{split}
B^{(j)}=&I_3+\frac{1}{\lambda}\left( \begin{array}{ccc}
a_j+\frac{i}{3}\Omega^{(j)} & \frac{i}{3}(u^{(j-1)}_1-u^{(j)}_1) & \frac{i}{3}(u^{(j-1)}_2-u^{(j)}_2) \\
\frac{i}{3}(\bar{u}^{(j-1)}_1-\bar{u}^{(j)}_1) &  a_j-ib_j-\frac{|u^{(j-1)}_1-u^{(j)}_1|^2}{3i(\Omega^{(j)}+3b_j)} & -\frac{(u^{(j-1)}_2-u^{(j)}_2)(\bar{u}^{(j-1)}_1-\bar{u}^{(j)}_1)}{3i(\Omega^{(j)}+3b_j)} \\
\frac{i}{3}(\bar{u}^{(j-1)}_2-\bar{u}^{(j)}_2) &  -\frac{(u^{(j-1)}_1-u^{(j)}_1)(\bar{u}^{(j-1)}_2-\bar{u}^{(j)}_2)}{3i(\Omega^{(j)}+3b_j)}  & a_j-ib_j-\frac{|u^{(j-1)}_2-u^{(j)}_2|^2}{3i(\Omega^{(j)}+3b_j)} \\
\end{array} \right)
\\\equiv &I_3+\frac{1}{\lambda}\hat{B}^{(j)},
\end{split}
\label{Bj}
\end{eqnarray}
where $\Omega^{(j)}=\sqrt{9b^2_j-|u^{(j-1)}_1-u^{(j)}_1|^2-|u^{(j-1)}_2-u^{(j)}_2|^2}$, $j=1,2$.
The corresponding BTs are
\begin{eqnarray}
\begin{split}
\left(u^{(j)}_{1}-u^{(j-1)}_{1}\right)_x=&u^{(j-1)}_{1}\left(\Omega^{(j)}-3ia_j\right)
+3iu^{(j)}_1\hat{B}^{(j)}_{22}+3iu^{(j)}_2\hat{B}^{(j)}_{32},
\\
\left(u^{(j)}_{2}-u^{(j-1)}_{2}\right)_x=&u^{(j-1)}_{2}\left(\Omega^{(j)}-3ia_j\right)
+3iu^{(j)}_1\hat{B}^{(j)}_{23}+3iu^{(j)}_2\hat{B}^{(j)}_{33},
\\
\left(u^{(j)}_{1}-u^{(j-1)}_{1}\right)_t=&i\left(u^{(j)}_{1}-u^{(j-1)}_{1}\right)
\left(|u^{(j)}_1|^2+|u^{(j)}_2|^2+|u^{(j-1)}_1|^2+|u^{(j-1)}_2|^2\right)
\\&+i\left(u^{(j-1)}_1u^{(j)}_2-u^{(j)}_1u^{(j-1)}_2\right)\bar{u}^{(j-1)}_2
  +iu^{(j-1)}_{1,x}\left(\Omega^{(j)}-3ia_j\right)
\\&-3u^{(j)}_{1,x}\hat{B}^{(j)}_{22}-3u^{(j)}_{2,x}\hat{B}^{(j)}_{32},
\\
\left(u^{(j)}_{2}-u^{(j-1)}_{2}\right)_t=&i\left(u^{(j)}_{2}-u^{(j-1)}_{2}\right)
\left(|u^{(j)}_1|^2+|u^{(j)}_2|^2+|u^{(j-1)}_1|^2+|u^{(j-1)}_2|^2\right)
\\&+i\left(u^{(j-1)}_2u^{(j)}_1-u^{(j)}_2u^{(j-1)}_1\right)\bar{u}^{(j-1)}_1
  +iu^{(j-1)}_{2,x}\left(\Omega^{(j)}-3ia_j\right)
\\&-3u^{(j)}_{1,x}\hat{B}^{(j)}_{23}-3u^{(j)}_{2,x}\hat{B}^{(j)}_{33},
\end{split}
\label{BTj}
\end{eqnarray}
where $\hat{B}^{(j)}_{km}$, $k,m=1,2,3$, are $km$-entries of the matrix $\hat{B}^{(j)}$ (see (\ref{Bj})).
With the above two BT matrices, we define a BT matrix $\hat{B}$ as $\hat{B}=B^{(2)}B^{(1)}$, which induces a BT from $u^{(0)}_1$ and $u^{(0)}_2$ to $u^{(2)}_1$ and $u^{(2)}_2$.
We impose Dirichlet boundary on the fields $u^{(1)}_1$ and $u^{(1)}_2$, that is
\begin{eqnarray}
\left.u^{(1)}_{1}(x,t)\right|_{x=0}=\left.u^{(1)}_{2}(x,t)\right|_{x=0}=0,
\label{DBu1}
\end{eqnarray}
and we consider the odd reductions between $(u^{(0)}_1, u^{(0)}_2)$ and $(u^{(2)}_1, u^{(2)}_2)$, that is
\begin{eqnarray}
u^{(0)}_{j}(x,t)=-u^{(2)}_{j}(-x,t),~~j=1,2.
\label{oddred}
\end{eqnarray}
We want to look for a relation between the parameters $(a_1,b_1)$ and $(a_2,b_2)$ such that the BTs (\ref{BTj}) evaluated at $x=0$ are compatible with the reduction (\ref{oddred}). After direct computations using (\ref{DBu1}), we find that the desired condition  is
\begin{eqnarray}
a_1=-a_2,~~b_1=b_2.
\label{reab12}
\end{eqnarray}
By using (\ref{DBu1}), (\ref{oddred}) and (\ref{reab12}), we obtain
\begin{eqnarray}
\left.\hat{B}\right|_{x=0}=I_3-\frac{a_2^2+b_2^2}{\lambda^2}I_3 +\frac{2i}{3\lambda}\left( \begin{array}{ccc}
\hat{\Omega}_{12}(0) & -u^{(2)}_1(0) & -u^{(2)}_2(0) \\
 -\bar{u}^{(2)}_1(0) &  -3b_2+\frac{|u^{(2)}_1(0)|^2}{\hat{\Omega}_{12}(0)+3b_2} & \frac{\bar{u}^{(2)}_1(0)u^{(2)}_2(0)}{\hat{\Omega}_{12}(0)+3b_2} \\
 -\bar{u}^{(2)}_2(0) & \frac{u^{(2)}_1(0)\bar{u}^{(2)}_2(0)}{\hat{\Omega}_{12}(0)+3b_2} & -3b_2+\frac{|u^{(2)}_2(0)|^2}{\hat{\Omega}_{12}(0)+3b_2} \\
 \end{array} \right),
\label{B0}
\end{eqnarray}
where $\hat{\Omega}_{12}(0)=\sqrt{9b_2^2-|u^{(2)}_1(0)|^2-|u^{(2)}_2(0)|^2}$.
In the following, we will use notations $a$ and $b$ instead of $a_2$ and $b_2$, and notations $u_j$ instead of $u^{(2)}_j$, $j=1,2$.
After applying proposition 2 to the BT matrix $\hat{B}$, we finally obtain that the $K(\lambda)$ matrix describing the boundary condition (\ref{bcase5}) are given by $K(\lambda)=\left.\hat{B}\right|_{x=0}$, that is
\begin{eqnarray}
\begin{split}
K(\lambda)&=\left(1-\frac{a^2+b^2}{\lambda^2}\right) I_3+\frac{2i}{3\lambda}\left( \begin{array}{ccc}
\Omega_{12}(0) & -u_1(0) & -u_2(0) \\
 -\bar{u}_1(0) &  -3b+\frac{|u_1(0)|^2}{\Omega_{12}(0)+3b} & \frac{\bar{u}_1(0)u_2(0)}{\Omega_{12}(0)+3b} \\
 -\bar{u}_2(0) & \frac{u_1(0)\bar{u}_2(0)}{\Omega_{12}(0)+3b} & -3b+\frac{|u_2(0)|^2}{\Omega_{12}(0)+3b} \\
 \end{array} \right)
 \\&\equiv \left(1-(a^2+b^2)\lambda^{-2}\right) I_3+\lambda^{-1} K^{(0)},
 \end{split}
 \label{K5}
\end{eqnarray}
where $\Omega_{12}(0)=\Omega_{12}(0,t)$ (recall that $\Omega_{12}(x,t)$ is defined by (\ref{omg12})).

In analogy to the $K(\lambda)$ matrices (\ref{K2}) and (\ref{K3}),
the $K(\lambda)$ matrix (\ref{K5}) is a non constant matrix,
we need to introduce suitable boundary Poisson brackets in order to calculate the Poisson brackets for (\ref{K5}).
As in the case of the $K(\lambda)$ matrix (\ref{K2}), we define the boundary Poisson brackets in the form of (\ref{PBBK2}), but with the quantities $K^{(0)}_{jk}$, $1 \leq j,k \leq 3$, defined by (\ref{K5}). Using this boundary Poisson brackets and the explicit expression (\ref{K5}) for the $K(\lambda)$ matrix, we can verify, after straightforward calculation, that (\ref{alga}) becomes an identity.
Thus integrability of the new boundary condition (\ref{bcase5}) is obtained.

\subsection{Conserved quantities for the boundary conditions}

We may derive the explicit forms of the conserved quantities by studying the large $\lambda$ asymptotic expansion of the trace of the monodromy matrix (\ref{gm}). However the calculations are much heavier (especially for the higher order conserved quantities). In this subsection we will derive the conserved quantities with a more convenient method which is based directly on the Lax pair formulation.

We first note that (\ref{algc}) can be interpreted as the compatibility condition of
\begin{subequations}
\begin{eqnarray}
\phi(0,t,\lambda)=K(\lambda)\phi(0,t,-\lambda),
 \label{blpx}
 \\
\phi_t(0,t,\lambda)=V(0,t,\lambda)\phi(0,t,\lambda).
 \label{blpt}
\end{eqnarray}
\label{BLP}
\end{subequations}
By virtue of the bulk Lax pair equations (\ref{LPxt}) and the boundary equations (\ref{BLP}), we find the following result, which is a generalisation of the analogous results in \cite{CCD2021} and \cite{CZ2012} in two ways: it generalises \cite{CCD2021} to the coupled case and it generalises \cite{CZ2012} to the time-dependent case.
\begin{proposition}
For the coupled NLS equation in the presence of a boundary condition resulting from (\ref{algc}), a generating function for the conserved quantities is given by
\begin{eqnarray}
\begin{split}
I(\lambda)=&\int_{0}^{\infty}\left[u_1(x,t)\left(\Gamma_1 (x,t,\lambda)-\Gamma_1 (x,t,-\lambda)\right)+
u_2(x,t)\left(\Gamma_2 (x,t,\lambda)-\Gamma_2 (x,t,-\lambda)\right)\right]dx
\\&+\ln\left(K_{11}(\lambda)+K_{12}(\lambda)\Gamma_1(0,t,-\lambda)+K_{13}(\lambda)\Gamma_2 (0,t,-\lambda)\right),
\end{split}
\label{GCD}
\end{eqnarray}
where $K_{jk}(\lambda)$, $1\leq j,k\leq3$, denote the $jk$-entries of the matrix $K(\lambda)$.
\end{proposition}
{\bf Proof}
We denote the first and the second terms on the right hand side of (\ref{GCD}) as $I_{bulk}$ and $I_{boundary}$, respectively. That is
\begin{subequations}
\begin{eqnarray}
I_{bulk}=\int_{0}^{\infty}\left[u_1(x,t)\left(\Gamma_1 (x,t,\lambda)-\Gamma_1 (x,t,-\lambda)\right)+
u_2(x,t)\left(\Gamma_2 (x,t,\lambda)-\Gamma_2 (x,t,-\lambda)\right)\right]dx,
\label{GCDa}
\\
I_{boundary}=\ln\left(K_{11}(\lambda)+K_{12}(\lambda)\Gamma_1(0,t,-\lambda)+K_{13}(\lambda)\Gamma_2 (0,t,-\lambda)\right).
\label{GCDb}
\end{eqnarray}
\label{GCDab}
\end{subequations}
By using (\ref{CL}), we obtain
\begin{eqnarray}
\begin{split}
-\frac{d I_{bulk}}{dt}=&V_{11}(0,t,\lambda)+V_{12}(0,t,\lambda)\Gamma_1(0,t,\lambda)+V_{13}(0,t,\lambda)\Gamma_2(0,t,\lambda)
\\&
-V_{11}(0,t,-\lambda)-V_{12}(0,t,-\lambda)\Gamma_1(0,t,-\lambda)-V_{13}(0,t,-\lambda)\Gamma_2(0,t,-\lambda).
\end{split}
\label{dIbulk}
\end{eqnarray}
This implies that in general, the bulk density (\ref{GCDa}) is no more conserved for the coupled NLS equation in the presence of a boundary condition. We will show below that the quantity (\ref{GCDb}) compensates exactly for the loss of conservation of the bulk density. From (\ref{blpx}), we obtain
\begin{subequations}
\begin{eqnarray}
\Gamma_1(0,t,\lambda)= \frac{K_{21}(\lambda)+K_{22}(\lambda)\Gamma_1(0,t,-\lambda)+K_{23}(\lambda)\Gamma_2(0,t,-\lambda)}
{K_{11}(\lambda)+K_{12}(\lambda)\Gamma_1(0,t,-\lambda)+K_{13}(\lambda)\Gamma_2(0,t,-\lambda)},
\\
\Gamma_2(0,t,\lambda)=\frac{K_{31}(\lambda)+K_{32}(\lambda)\Gamma_1(0,t,-\lambda)+K_{33}(\lambda)\Gamma_2(0,t,-\lambda)}
{K_{11}(\lambda)+K_{12}(\lambda)\Gamma_1(0,t,-\lambda)+K_{13}(\lambda)\Gamma_2(0,t,-\lambda)}.
\end{eqnarray}
\label{Gam1}
\end{subequations}
Equation (\ref{algc}) implies
\begin{eqnarray}
\begin{split}
\frac{d K_{1j}(\lambda)}{dt}=
&V_{11}(0,t,\lambda)K_{1j}(\lambda)+V_{12}(0,t,\lambda)K_{2j}(\lambda)+V_{13}(0,t,\lambda)K_{3j}(\lambda)
\\&
-K_{11}(\lambda)V_{1j}(0,t,-\lambda)-K_{12}(\lambda)V_{2j}(0,t,-\lambda)-K_{13}(\lambda)V_{3j}(0,t,-\lambda),
~1\leq j\leq 3.
\end{split}
\label{K123}
\end{eqnarray}
Inserting (\ref{Gam1}) into the right hand side of (\ref{dIbulk}), we find that the resulting expression is equal to
\begin{eqnarray}
\frac{\left(K_{11}(\lambda)+K_{12}(\lambda)\Gamma_1(0,t,-\lambda)+K_{13}(\lambda)\Gamma_2 (0,t,-\lambda)\right)_t}{K_{11}(\lambda)+K_{12}(\lambda)\Gamma_1(0,t,-\lambda)+K_{13}(\lambda)\Gamma_2 (0,t,-\lambda)},
\label{Ibd}
\end{eqnarray}
where we have used the formulae (\ref{K123}) and the Riccati equations (\ref{rict}) for $\Gamma_1(0,t,-\lambda)$ and $\Gamma_2(0,t,-\lambda)$.
The expression (\ref{Ibd}) is exactly the time derivative of (\ref{GCDb}). As a consequence, we obtain $\frac{dI(\lambda)}{dt}=0$. This completes the proof.
\QEDB

Based on the above proposition and the $K(\lambda)$ matrices found in section 5.3, we immediately obtain the following
corollaries.
\begin{corollary}\label{corollary1}
A generating function for the conserved quantities of the coupled NLS equation (\ref{CNLS}) with the Robin boundary condition (\ref{RBC}) is given by
\begin{eqnarray}
\begin{split}
I(\lambda)=&\int_{0}^{\infty}\left[u_1(x,t)\left(\Gamma_1 (x,t,\lambda)-\Gamma_1 (x,t,-\lambda)\right)+
u_2(x,t)\left(\Gamma_2 (x,t,\lambda)-\Gamma_2 (x,t,-\lambda)\right)\right]dx.
\end{split}
\label{GCDcase1}
\end{eqnarray}
\end{corollary}

\begin{corollary}\label{corollary2}
A generating function for the conserved quantities of the coupled NLS equation (\ref{CNLS}) with the new boundary condition (\ref{tbc}) is given by
\begin{eqnarray}
\begin{split}
I(\lambda)=&\int_{0}^{\infty}\left[u_1(x,t)\left(\Gamma_1 (x,t,\lambda)-\Gamma_1 (x,t,-\lambda)\right)+
u_2(x,t)\left(\Gamma_2 (x,t,\lambda)-\Gamma_2 (x,t,-\lambda)\right)\right]dx
\\&+\ln\left(1+\frac{i}{3}\lambda^{-1}\left(\hat{\Omega}(0,t)-2u_1(0,t)\Gamma_1 (0,t,-\lambda)
-2u_2(0,t)\Gamma_2 (0,t,-\lambda)\right)\right).
\end{split}
\label{GCDcase2}
\end{eqnarray}
\end{corollary}

\begin{corollary}\label{corollary3}
A generating functions for the conserved quantities of the coupled NLS equation (\ref{CNLS}) with the new boundary condition (\ref{bcase3}) is given by
\begin{eqnarray}
\begin{split}
I(\lambda)=&\int_{0}^{\infty}\left[u_1(x,t)\left(\Gamma_1 (x,t,\lambda)-\Gamma_1 (x,t,-\lambda)\right)+
u_2(x,t)\left(\Gamma_2 (x,t,\lambda)-\Gamma_2 (x,t,-\lambda)\right)\right]dx
\\&+\ln\left(1+\frac{i}{3}\lambda^{-1}\left(\Omega_{2}(0,t)-2u_2(0,t)\Gamma_2 (0,t,-\lambda)\right)\right).
\end{split}
\label{GCDcase3}
\end{eqnarray}
\end{corollary}

\begin{corollary}\label{corollary4}
A generating function for the conserved quantities of the coupled NLS equation (\ref{CNLS}) with the new boundary condition (\ref{bcase5}) is given by
\begin{eqnarray}
\begin{split}
I(\lambda)=&\int_{0}^{\infty}\left[u_1(x,t)\left(\Gamma_1 (x,t,\lambda)-\Gamma_1 (x,t,-\lambda)\right)+
u_2(x,t)\left(\Gamma_2 (x,t,\lambda)-\Gamma_2 (x,t,-\lambda)\right)\right]dx
\\&+\ln\left(1-\frac{a^2+b^2}{\lambda^2}+\frac{2i}{3\lambda}\left(\Omega_{12}(0,t)-u_1(0,t)\Gamma_1(0,t,-\lambda)-u_2(0,t)\Gamma_2 (0,t,-\lambda)\right)\right).
\end{split}
\label{GCDcase5}
\end{eqnarray}
\end{corollary}

By substituting (\ref{Gamexpansion}) together with (\ref{wj}) into the bulk density (\ref{GCDa}) and by considering the series expansion of the boundary contribution (\ref{GCDb}) in $\lambda^{-1}$, we can compute the explicit forms of the conserved quantities for our boundary problems.
For example, the first two conserved quantities for the coupled NLS equation with the new boundary condition (\ref{tbc}) are given by:
\begin{eqnarray}
\begin{split}
I_1=&-2\int_0^{\infty}\left(|u_1|^2+|u_2|^2\right)dx+\left.\hat{\Omega}\right|_{x=0},
\\
I_3=&\int_0^{\infty}\left(\left(|u_1|^2+|u_2|^2\right)^2-|u_{1,x}|^2-|u_{2,x}|^2\right)dx
-\left.\left(\frac{1}{6}\hat{\Omega}^3+\hat{\Omega}\left(|u_1|^2+|u_2|^2\right)\right)\right|_{x=0},
\end{split}
\label{ecd}
\end{eqnarray}
where $\hat{\Omega}$ is given by (\ref{hatomega}).

\section{Concluding remarks}

Based on a BT of the coupled NLS equation, we presented a type I defect condition and established the Liouville integrability of the resulting defect system. Furthermore, by imposing suitable reductions on the defect condition,
we derived several new integrable boundary conditions for the coupled NLS equation. A remarkable feature of these new integrable boundary conditions is the presence of time derivatives of the coupled NLS fields.
The boundary matrices that realise our new boundary conditions were also derived by virtue of a connection between the time-part of the BT equations and the Sklyanin's formalism.

We end this paper with the following remarks.

1. The coupled NLS equation can be further generalised to the multi-component case, the so-called vector NLS model  \cite{Faddeev2007}. Based on our experience with the coupled NLS equation, we believe that it is possible to work out the type I defect conditions and the associated new integrable boundary conditions for the vector NLS model (these problems were identified as open problems in \cite{Zambon2014}). We will study this topic in the future.

2. Very recently, the studies on the solutions of the NLS equation in the presence of the boundary condition (\ref{tbcNLS}) were performed in \cite{Xia2021,Gruner2020,Zhangc2021,CCD2021}.
As mentioned previously, our new integrable boundary conditions are two-component analogues of the integrable boundary condition (\ref{tbcNLS}). It will be interesting to study whether the solution methods developed in \cite{Xia2021,Gruner2020,Zhangc2021,CCD2021} can be extended to study our new integrable boundary conditions for the coupled NLS equation. We strongly believe that this is the case, despite additional computational and technical difficulties that will arise in our situation.


\section*{ACKNOWLEDGMENTS}
The author would like to express his sincerest thanks to the referees for their valuable comments, which have helped him  to improve this paper.
This work was supported by the National Natural Science Foundation of China (Grant No. 11771186) and 333 Project of Jiangsu Province (No. BRA2020246).

\vspace{1cm}
\small{

}

\begin{thebibliography}{99}

\bibitem{Manakov} S.V. Manakov, On the theory of two-dimensional stationary self focussing of
electromagnetic waves, Soviet Phys. JETP 38 (1974) 248-253.
\bibitem{BuAn} T. Busch and J.R. Anglin, Dark-bright solitons in inhomogeneous Bose-Einstein condensates, Phys. Rev. Lett. 87 (2001) 010401.

\bibitem{BCZ20041} P. Bowcock, E. Corrigan and C. Zambon, Classically integrable field theories with defects, Int. J. Mod. Phys. A 19 (2004) 82-91;
    \\P. Bowcock, E. Corrigan and C. Zambon, Affine Toda field theories with defects, J. High Energ. Phys. 2004 (2004) 056.
\bibitem{CZ2006} E. Corrigan and C. Zambon, Jump-defects in the nonlinear Schr\"{o}dinger model and other non-relativistic field theories, Nonlinearity 19 (2006) 1447.
\bibitem{CZ2009} E. Corrigan and C. Zambon, A new class of integrable defects, J. Phys. A 42 (2009) 475203.

\bibitem{HK2008} I. Habibullin and A. Kundu, Quantum and classical integrable sine-Gordon model with defect, Nucl. Phys. B 795 (2008) 549.
\bibitem{Caudrelier2008} V. Caudrelier, On a systematic approach to defects in classical integrable field theories, Int. J. Geom. Meth. Mod. Phys. vol. 5, No. 7 (2008) 1085.
\bibitem{CK2015} V. Caudrelier and A. Kundu, A multisymplectic approach to defects in integrable classical field theory, J. High Energ. Phys. 02 (2015) 088.
\bibitem{AD2012} J. Avan and  A. Doikou, Liouville integrable defects: the non-linear Schr\"{o}dinger paradigm, J. High Energ. Phys.  2012 (2012) 40.
\bibitem{Doikou2016} A. Doikou, Classical integrable defects as quasi B\"{a}cklund transformations, Nucl. Phys. B 911 (2016) 212.

\bibitem{Zambon2014} C. Zambon, The classical nonlinear Schr\"{o}dinger model with a new integrable boundary, J. High Energ. Phys. 2014 (2014) 36.
\bibitem{Doikou2014} A. Doikou, Classical impurities associated to high rank algebras, Nucl. Phys. B 884 (2014) 142.


\bibitem{ACDK2016} J. Avan, V. Caudrelier, A. Doikou, and A. Kundu, Lagrangian and Hamiltonian structures in an integrable hierarchy and space-time duality, Nucl. Phys. B 902 (2016) 415-439.

\bibitem{Gruner2020} K.T. Gruner, Dressing a new integrable boundary of the nonlinear Schr\"{o}dinger equation, arXiv preprint arXiv:2008.03272;
    \\K. T. Gruner, Soliton solutions of the nonlinear Schr\"{o}dinger equation with defect conditions, Nonlinearity, 34 (2021) 6017. 
\bibitem{Xia2021} B. Xia, On the nonlinear Schr\"{o}dinger equation with a boundary condition involving a time
derivative of the field, J. Phys. A: Math. Theor. 54 (2021) 165202.
\bibitem{Zhangc2021} C. Zhang. On the inverse scattering transform for the nonlinear Schr\"{o}dinger equation on the half-line, arXiv preprint arXiv:2106.02336, 2021.

\bibitem{CCD2021} V. Caudrelier, N. Crampe and C.M. Dibaya, Nonlinear mirror image method for nonlinear Schr\"{o}dinger equation: Absorption/emission of one soliton by a boundary, Stud. Appl. Math. 148 (2022) 715-757. 


\bibitem{Sklyanin1987} E.K. Sklyanin, Boundary conditions for integrable equations. Functional Analysis and its Applications, 21(2) (1987) 164-166.

\bibitem{zhou2017} R. Zhou, P. Li and Y. Gao, Equal-Time and Equal-Space Poisson Brackets of the N-Component Coupled NLS Equation, Commun. Theor. Phys. 67 (2017) 347-349.

\bibitem{Faddeev2007} L.D. Faddeev, L.A. Takhtajan, Hamiltonian Methods in the Theory of Solitons, Springer (2007).

\bibitem{Caudrelier2020} V. Caudrelier, M. Stoppato, A connection between the classical $r$-matrix formalism
and covariant Hamiltonian field theory, J. Geom. Phys. 148 (2020) 103546.

\bibitem{DFR} A. Doikou, D. Fioravanti and F. Ravanini, The generalized non-linear Schr\"{o}dinger model on the interval, Nuclear Physics B 790 (2008) 465-492.

\bibitem{Wright2003} O.C. Wright, The Darboux Transformation of Some Manakov Systems, Applied Mathematics Letters, 16 (2003) 647-652.
\bibitem{Gu} C.H. Gu, H.S. Hu and Z.X. Zhou, Darboux Transformations in Integrable Systems: Theory and Their Applications (Berlin: Springer) 2005.

\bibitem{KW1976} Y. Kodama and M. Wadati, Theory of canonical transformations for nonlinear evolution equations
.I, Prog. Theo. Phys. 56 (1976) 1740-1755.
\bibitem{K1977} Y. Kodama, Theory of canonical transformations for nonlinear evolution equations. II,
Prog. Theo. Phys. 57 (1977) 1900-1916.

\bibitem{Habibullin1991} I.T. Habibullin, The B\"{a}cklund transformation and integrable initial boundary value problems. Matematicheskie Zametki, 49(4) (1991) 130-137.

\bibitem{CZ2012} V. Caudrelier and C. Zhang, The vector nonlinear Schr\"{o}dinger equation on the half-line, Journal of Physics A: Mathematical and Theoretical, 45(10) (2012) 105201.
\bibitem{ZZ2021} C. Zhang and D. Zhang, Vector NLS solitons interacting with a boundary, Communications in
Theoretical Physics, 73(4) (2021) 045005.

\bibitem{ACC2018} J. Avan, V. Caudrelier and N. Cramp\'{e}, From Hamiltonian to zero curvature formulation for classical
integrable boundary conditions, J. Phys. A 51 (2018) 30LT01.














\end{thebibliography}
\end{document}